\documentclass[aps,prb,groupedaddress,showpacs,twocolumn]{revtex4}
\usepackage{graphicx}
\usepackage{amsmath}
\usepackage{amssymb}
\usepackage{bbm}
\usepackage{xspace}
\usepackage{color}
\usepackage{footnote}

\definecolor{orange}{RGB}{252,77,6}
\definecolor{brown}{RGB}{200,127,50}
\definecolor{green1}{RGB}{00,100,00}
\definecolor{green2}{RGB}{00,150,00}
\definecolor{green3}{RGB}{00,200,00}
\definecolor{green4}{RGB}{00,250,00}
\definecolor{gray}{RGB}{150,150,150}

\newcommand{\cdag}[1]{c^{\dagger}_{#1}}
\newcommand{\cnodag}[1]{c^{\phantom{\dagger}}_{#1}}

\newcommand{\fig}[1]{Fig.\thinspace{}\ref{#1}}

\newcommand{\Fig}[1]{Fig.\thinspace{}\ref{#1}}
\newcommand{\eq}[1]{Eq.\thinspace{}(\ref{#1})}

\newcommand{\se}{Sec.\@\xspace}

\newcommand{\app}{App.\@\xspace}
\newcommand{\App}{App.\@\xspace}
\newcommand{\thickbar}[1]{\mathbf{\bar{\text{$#1$}}}}

\newcommand{\etal}[0]{\textit{et al.}}
\newcommand{\tcite}[1]{Ref.~\onlinecite{#1}}
\newcommand{\tcites}[1]{Refs.~\onlinecite{#1}}

\def\ket#1{\mathinner{|{#1}\rangle}}

\newcommand{\nag}{{\phantom{\dag}}}

\newcommand{\ve}[1]{\boldsymbol{\mathbf{#1}}}

\usepackage{hyperref}

\begin{document}


\title{Nonequilibrium, spatiotemporal formation of the Kondo screening cloud on a lattice}


\author{Martin Nuss}
\email[]{martin.nuss@tugraz.at}
\affiliation{Institute of Theoretical and Computational Physics, Graz University of Technology, 8010 Graz, Austria}
\author{Martin Ganahl}
\affiliation{Institute of Theoretical and Computational Physics, Graz University of Technology, 8010 Graz, Austria}
\author{Enrico Arrigoni}
\affiliation{Institute of Theoretical and Computational Physics, Graz University of Technology, 8010 Graz, Austria}
\author{Wolfgang von der Linden}
\affiliation{Institute of Theoretical and Computational Physics, Graz University of Technology, 8010 Graz, Austria}
\author{Hans Gerd Evertz}
\affiliation{Institute of Theoretical and Computational Physics, Graz University of Technology, 8010 Graz, Austria}

\date{\today}

\begin{abstract}
We study the nonequilibrium formation of a spin screening cloud that accompanies the quenching of a local magnetic moment immersed in a Fermi sea at zero temperature. Based on high precision density matrix renormalization group results for the interacting single impurity Anderson model we discuss the real time evolution after a quantum quench in the impurity-reservoir hybridization using time evolving block decimation. We report emergent length and time scales in the spatiotemporal structure of non-local correlation functions in the spin- and the charge density channel. At equilibrium, our data for the correlation functions and the extracted length-scales show good agreement with existing results, as do local time dependent observables at the impurity. In the time-dependent data, we identify a major signal which defines a ``light cone'' moving at the Fermi velocity and a ferromagnetic component in its wake. Inside the light cone we find that the structure of the nonequilibrium correlation functions emerges on two time scales. Initially, the qualitative structure of the correlation functions develops rapidly at the lattice Fermi velocity. Subsequently the spin correlations converge to the equilibrium results on a much larger time scale. This process sets a dynamic energy scale, which we identify to be proportional to the Kondo temperature. Outside the light cone we observe two different power law decays of the correlation functions in space, with time and interaction strength independent exponents.
\end{abstract}

\pacs{72.10.Fk, 72.15.Qm, 71.27.+a, 73.21.La}

\maketitle

\section{Introduction}\label{sec:modelAndMethod}
Quantum impurities are among the most fundamental paradigms of strongly correlated quantum systems. Equilibrium properties of such systems have been subject to intense investigations and are nowadays well understood. A famous example is the Kondo effect, where a local spin-$\frac{1}{2}$ degree of freedom interacts with the spins of a sea of free electrons.~\cite{hews.97} The ground state of this system is a delocalized spin singlet, formed by the local moment and the spin of the free electrons, also called a screening cloud. The present work investigates how such a screening cloud develops over time when a local moment comes into contact with a free electron reservoir.

Quantum impurity systems, quite generally, feature an emergent screening length-scale at low temperatures which provides the basis for their complex physics. In the 1950s, magnetic impurities have already been identified as the cause for a large resistivity anomaly at low temperatures when immersed in metallic hosts in dilute quantities.~\cite{frie.56, cl.ma.62} It was found theoretically that the impurity's local magnetic moment becomes quenched below a certain temperature, known as the Kondo temperature~\cite{kond.64,hews.97} $T_{\text{K}}$, to form a local Fermi liquid.~\cite{nozi.74} Increased spin flip scattering between pairs of degenerate spin-$\frac{1}{2}$ states then leads to an increase in resistivity below $T_{\text{K}}$.

Meanwhile, the Kondo effect has been observed also in nanoscopic devices like quantum dots,~\cite{go.go.98,cr.oo.98,si.bl.99,wi.fr.00,wi.fr.02,fr.ha.02,kr.sh.12} carbon nano tubes~\cite{ch.ga.12} and molecular junctions.~\cite{li.sh.02} Here, the narrow, zero energy, resonance in the local density of states of the impurity, the Kondo-Abrikosov-Suhl resonance, leads to a well defined unitary conductance in linear response. The Kondo effect has also proven essential to understanding tunnelling into single magnetic atoms,~\cite{ma.ch.98} adsorption of molecules onto surfaces~\cite{br.sc.74} or defects in materials such as graphene.~\cite{ch.li.11} On the theoretical side the Kondo effect lies at the heart of our current understanding of correlated materials, notably within the very successful dynamical mean-field theory (DMFT).~\cite{ge.ko.96,voll.10,me.vo.89}

Insight into the details of the screening cloud is important not only for the  understanding of the physics of a single impurity but also for the understanding of the interplay of many magnetic impurities. Many impurities result in competing effects among conduction electrons and local moments which form the basis for spin exhaustion scenarios~\cite{nozi.85,nozi.98} as well as for the Doniach phase diagram~\cite{doni.77,doni.77b} which describes the relationship between Kondo~\cite{hews.97} and RKKY interaction.~\cite{ru.ki.54,kasu.56,yosi.57}

Experimental characterization of the structure of the singlet ground state, which is a bound state of the impurity spin and the reservoir electron "screening cloud", has proven difficult so far. Several proposals exist for how to measure the spatial extent of the spin screening cloud or its antiferromagnetic correlation with the impurity spin.~\cite{affl.09,affl.01} In principle, the real space structure could be probed by performing nuclear magnetic resonance (NMR) / Knight shift~\cite{knig.49,to.he.50,knig.56} measurements on bulk metals hosting dilute magnetic impurities but the approach remains challenging.~\cite{affl.09} Indirect observation by measurement of the Kondo resonance, for example by photo emission also remains elusive due to the too narrow resonance at the Fermi energy.~\cite{ha.kr.06} Other proposals suggest the use of scanning tunnelling microscopy (STM)~\cite{bi.ro.86} and scanning tunnelling spectroscopy (STS) to analyse adatoms or surface defects with Kondo behaviour.~\cite{pr.wa.11,af.bo.08} In the realm of nano devices, proposals include experiments based on persistent currents~\cite{af.pa.01} or in confined geometries.~\cite{pe.la.08,pa.le.13} Some progress has been made recently using single magnetic atoms~\cite{ma.ch.98}, quantum corrals~\cite{ma.lu.00} or impurities beneath surfaces.~\cite{pr.wa.11}

On the theoretical side, however, the structure of the screening cloud 
has been characterized, at least in the equilibrium spin-spin correlation function~\cite{iish.78} and the charge density-density correlation function.~\cite{me.gr.72,zl.gr.77} Theoretical results~\cite{ba.af.96,so.af.96,so.af.05,co.ga.06} at equilibrium include studies employing quantum Monte Carlo (QMC),~\cite{gu.hi.87} numerical renormalization group (NRG)~\cite{bord.07,bo.ga.09,bu.ma.10} as well as density matrix RG (DMRG)~\cite{ho.hm.09} and in the noninteracting system also exact diagonalization (ED).~\cite{go.ri.14}

\begin{figure}
\includegraphics[width=0.4\textwidth]{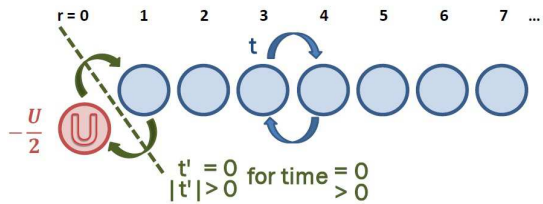}
\caption{(Color online) \emph{The model} consists of a fermionic impurity with local Coulomb repulsion $U$, which is coupled to a one-dimensional half-filled tight-binding chain in a particle-hole symmetric fashion. At time~\cite{footnote3} $\tau=0$ we switch on the tunnelling $t'$ and study the evolution  of the spin- and charge correlation functions: $S(r,\tau)$ and $C(r,\tau)$.}
\label{fig:fig0a}
\end{figure}

The fascinating question of how the spatial structure of the Kondo screening cloud develops in the first place, i.e. whether and how it is approached in a nonequilibrium time evolution starting from an initial state without Kondo physics has recently come under investigation.~\cite{hoff.12,me.ho.13, le.an.14} The present work extends previous equilibrium DMRG calculations~\cite{ho.hm.09} by investigating the dynamic nonequilibrium formation of Kondo correlations. We study the physical behaviour of the single impurity Anderson model (SIAM)~\cite{ande.61} based on results obtained with DMRG~\cite{whit.92,whit.93,scho.05} and the time evolving block decimation (TEBD)~\cite{vida.04} for matrix product states (MPS).~\cite{scho.11} The system is sketched in \fig{fig:fig0a}. At time~\cite{footnote3} $\tau=0$ we start from an unentangled state of a singly occupied impurity and a half filled Fermi sea (FS) of conduction electrons \hbox{$\ket{\Psi}=\ket{\uparrow}_{\text{impurity}}\otimes\ket{\text{FS}}_{\text{reservoir}}$}. Then, after connecting the impurity to the reservoir we follow the evolution of correlation functions over time as the system equilibrates and the "impurity spin gets transported to infinity". In this way, we obtain information about the spatiotemporal structure of the screening cloud.

Recently, studies of the time-dependent behaviour of length-scales in strongly correlated impurity systems were performed for the Toulouse point of the anisotropic Kondo model, where it maps onto a noninteracting system~\cite{hoff.12,me.ho.13} and for the symmetric Kondo model.~\cite{le.an.14} In both these systems a ``light cone'' like propagation of excitations with the Fermi velocity was observed and the regions inside as well as outside the light cone were investigated. Both studies identified a common low energy scale: the inverse Kondo temperature as a time scale, which was seen in a spin correlation function outside the light cone at the Toulouse point.~\cite{hoff.12,me.ho.13} For the symmetric Kondo model such a time scale was observed in an equilibrium linear response calculation to a magnetic perturbation.~\cite{le.an.14} We extend these studies to the SIAM which shares a common low energy behaviour with the symmetric Kondo model. Whereas in the Kondo model only spin interactions survive and charge fluctuations are treated on an effective level,~\cite{sc.wo.66} we take them into account explicitly. To our knowledge our study is the first one analysing the nonequilibrium properties of the screening length in the interacting SIAM. 

\begin{figure}
\includegraphics[width=0.49\textwidth]{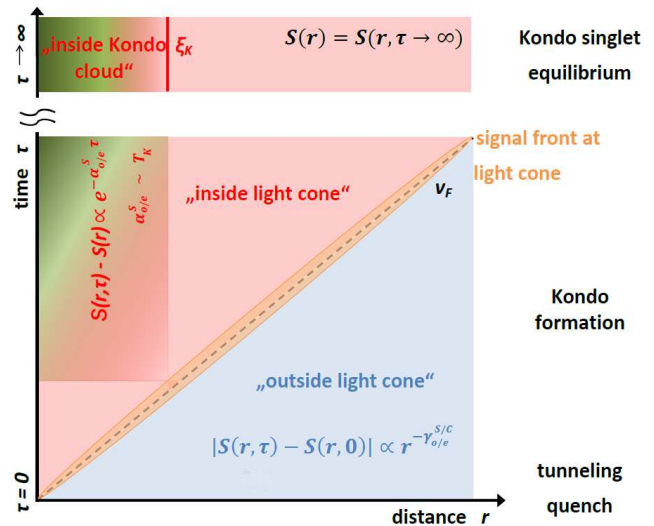}
\caption{(Color online) \emph{Schematic summary of results of this paper.} The time evolution of the spin correlation function $S(r,\tau)$ exhibits three characteristic regions in space and time. These are divided by i) a major signal following the quench, which propagates at the lattice Fermi velocity $v_{\text{F}}$ and defines a light cone (dashed line) and ii) the spread of the Kondo spin screening cloud. Region one (green) lies inside the light cone and inside the Kondo screening cloud. Here Kondo correlations develop on two characteristic time scales. The main structure of the Kondo singlet correlations is formed rapidly at $v_{\text{F}}$. Then these correlations approach their equilibrium values exponentially slowly in time for $\tau\rightarrow\infty$, 
with an exponent $\alpha^S_{\rm{o/e}}$ that is proportional to the Kondo temperature $T_{\text{K}}$. Region two (red) lies inside the light cone but outside the Kondo screening cloud. Here the spin correlations decay as a power law in space.~\cite{go.ri.14} In region three (blue), which lies outside the light cone and outside the Kondo screening cloud, the correlation function at odd/even distances decays as a power law $\propto r^{-\gamma^{S/C}_{\rm{o/e}}}$ in space with exponents that are independent of time and interaction strength.}
\label{fig:fig0b}
\end{figure}

Our results are summarized in \fig{fig:fig0b} which also serves as a guiding map for this work. We identify a major signal following the quench, which propagates at the lattice Fermi velocity $v_{\text{F}}$ and defines a light cone for the propagation of information.~\cite{li.ro.72,br.ha.06,na.si.06,sc.ha.11,hast.12, hoff.12,me.ho.13,le.an.14} Inside the light cone the time evolved correlation functions converge to their equilibrium counterparts which exhibit the Kondo length-scale. We find that Kondo correlations develop on two characteristic time scales. The main structure of the Kondo singlet is formed rapidly at $v_{\text{F}}$. These correlations approach their equilibrium values exponentially in time, defining a dynamic energy scale $\alpha^{S}_{\rm{o/e}}$, which is proportional to the Kondo temperature $T_{\text{K}}$. Outside the light cone, we find that correlation functions at odd/even distances decay as a power law $\propto r^{-\gamma^{S/C}_{\rm{o/e}}}$ in space, with exponents which are independent of time and interaction strength.

The structure of this paper is as follows: We summarize the specific model used in \se~\ref{sec:Model}. We define the Kondo singlet in \se~\ref{sec:KondoSinglet}, present our numerical approach in \se~\ref{sec:Method} and provide an overview of the equilibrium situation in \se~\ref{sec:equilibrium}. We start our presentation of nonequilibrium phenomena in \se~\ref{sec:localObs}, where we discuss the evolution of local observables. The main findings of this work are available in \se~\ref{sec:nonequilibrium}. There we discuss the nonequilibrium formation of the Kondo screening cloud in \se~\ref{sec:inside}. The situation outside the light cone is presented in \se~\ref{sec:outside}. The quality of our numerical data is assessed in \app~\ref{app:NumericalDetails}.

\section{Model}\label{sec:Model}
We study a lattice realization of the SIAM~\cite{ande.61}
\begin{align}
  \hat{\mathcal{H}}_{\text{SIAM}} &= \hat{\mathcal{H}}_{\text{imp}} + \hat{\mathcal{H}}_{\text{tunn}} + \hat{\mathcal{H}}_{\text{res}}\,\mbox{,}
\label{eq:HSIAM}
\end{align}
which consists of a single fermionic spin-$\frac{1}{2}$ impurity coupled via a standard hopping term to a reservoir of noninteracting tight-binding fermions, see \fig{fig:fig0a}. In particular we consider a particle-hole symmetric impurity with on-site interaction $U$
\begin{align}
  \hat{\mathcal{H}}_{\text{imp}} &= -\frac{U}{2} \, \sum\limits_{\sigma}  \, f_{\sigma}^\dag \, f_{\sigma}^\nag + U \, \hat{n}_{f \uparrow} \, \hat{n}_{f \downarrow}\,\mbox{.}
\end{align}
The electronic annihilation (creation) operators $f_{\sigma}^{\nag}$ ($f_{\sigma}^{\dag}$) obey the usual anti-commutation relations with spin $\sigma = \{\uparrow, \downarrow\}$, and $\hat{n}_{f \sigma}=f_{\sigma}^\dagger \, f_{\sigma}^\nag$ is the particle number operator.~\cite{footnote10} The impurity is coupled via a tunnelling term
\begin{align*}
\hat{\mathcal{H}}_{\text{tunn}} &= -t' \sum\limits_{\sigma}\left(c_{1\sigma}^{\dagger} \, f_{\sigma}^\nag + f_{\sigma}^\dagger \, c_{1\sigma}^{\nag} \right)\,\mbox{,}
\end{align*}
to a one-dimensional tight-binding chain
\begin{align*}
\hat{\mathcal{H}}_{\text{res}} &= - t\sum\limits_{\sigma}\,  \, \sum\limits_{i=1}^{L-2}  \, \left(c_{i\sigma}^{\dagger} \, c_{i+1\sigma}^{\nag} + c_{i+1\sigma}^{\dagger} \, c_{i\sigma}^{\nag}\right)\,\mbox{,}
\end{align*}
such that the overall system, including the impurity is of length $L$. We always take the reservoir Fermi sea $\hat{\mathcal{H}}_{\text{res}}$ half-filled. For large $L$, the reservoir mimics a semi-infinite one-dimensional tight-binding reservoir~\cite{econ.10} with semi-circular density of states at the first site and bandwidth $D = 4\,t$.~\cite{footnote7} Studies of finite size effects are available in \tcite{so.af.96,th.kr.99,af.pa.01,pa.af.03,ha.kr.06,sc.gu.12,ti.po.13}. The hopping parameter of the reservoir $t$ is taken to be unity, and its coupling to the impurity $t'=0.3162\,t$ combine to an equilibrium Anderson width~\cite{hews.97} of $\Delta \equiv \pi\,t'^2 \,\rho_{\text{reservoir}}(0) = \frac{t'^2}{t} \approx 0.1\,t$, where $\rho_{\text{reservoir}}(\omega)$ denotes the density of states of the reservoir.

At equilibrium, many characteristics of the SIAM are known although it poses a difficult interacting problem. Seminal results for the ground state and thermodynamic properties of the SIAM at equilibrium are available from perturbation theory (PT),~\cite{yo.ya.70,yama.75a,yo.ya.75,yama.75b} RG~\cite{ande.70,wils.75,kr.wi.80,bu.co.08} and the Bethe Ansatz (BA).~\cite{beth.31,wi.ts.83a,ts.wi.83b} Hirsch-Fye QMC~\cite{hi.fy.86,gu.hi.87} and the continuous time QMC~\cite{gu.mi.11} accurately describe the imaginary time dynamics. Further, some physical results can be inferred from the Kondo Hamiltonian, which is related to the SIAM by the Schrieffer-Wolff transformation to obtain its low energy realization, in which charge fluctuations are integrated out.~\cite{sc.wo.66,ande.70} 

\section{Kondo singlet}\label{sec:KondoSinglet}
At equilibrium, the SIAM features a characteristic length scale which, for finite interaction strength, is the Kondo length-scale and is expected to correspond  to the size of the singlet screening cloud. This length-scale is defined as~\cite{hald.78,affl.09,iish.78,ba.af.98,zi.bo.06,bord.07} $\xi_{\text{K}}\equiv \frac{v_{\text{F}}}{T_{\text{K}}}$, i.e. it is proportional to the Fermi velocity $v_{\text{F}}\approx 2t$ and the inverse Kondo temperature $\frac{1}{T_{\text{K}}}$.~\cite{wi.ts.83a,ts.wi.83b,hews.97} $T_{\text{K}}$ can be extracted from many observables, most intuitive is the definition as the temperature at which the local moment becomes quenched, i.e. when the impurity entropy goes from $\text{ln}(2)$, indicating the local moment regime, to $\text{ln}(1)$, indicating the singlet state.~\cite{prus.12} A scale proportional to $T_{\text{K}}$ is also available from the zero temperature self-energy~\cite{ka.he.08} or from the width of the Kondo resonance in the spectral function.~\cite{cole.02} An analytic expression for $T_{\text{K}}$, as obtained from the spin susceptibility, is available for the SIAM at particle-hole symmetry in the wide band limit with a linear dispersion~\cite{footnote8} by the BA:~\cite{beth.31,wi.ts.83a,ts.wi.83b} $T_{\text{K}}^{\text{BA}}=\sqrt{\Delta U}e^{-\frac{\pi}{16 \Delta}U}$. The Kondo singlet, therefore, is exponentially large in the interaction strength $U$
\begin{align}
  \xi_{\text{K}}^{\text{BA}}&\approx\frac{2t}{\sqrt{\Delta U}}e^{\frac{\pi}{16 \Delta}U}\,\mbox{.}
\label{eq:xiKBA}
\end{align}
For typical Kondo materials, like dilute magnetic impurities in free electron metals~\cite{as.me.76} one finds $v_{\text{F}}\approx10^{6}\,\frac{\text{m}}{\text{s}}$ and $T_{\text{K}}\approx1\,$K valid for example in gold with dilute iron impurities.~\cite{ba.ko.78} Thus, the screening length becomes macroscopic $\xi_{\text{K}}\approx1\,\mu$m.~\cite{bord.07}

Here, we extract the screening length-scale $\xi_{\text{K}}$ directly from correlation functions and not via the Kondo temperature. The spin correlation function is defined as
\begin{align}
S(r,\tau) &= \langle\hat{\ve{S}}_0\cdot\hat{\ve{S}}_r\rangle(\tau)\,\mbox{,}
\label{eq:S}
\end{align}
where $\hat{\ve{S}}_r = (\hat{S}^x_r,\hat{S}^y_r,\hat{S}^z_r)$,~\cite{footnote11}
and  $r$ denotes the distance from the impurity in units of the lattice spacing (see \fig{fig:fig0a}).
Due to the oscillations of $S$, it is convenient to  distinguish between the spin correlation function for odd  ($S_{\rm{o}}(r,\tau)$) and that for even ($S_{\rm{e}}(r,\tau)$) distances.

Length scales can be extracted from the crossover in the functional dependence of $S_{\rm{o}}(r,\tau)$ or via determining zeros or minima in $S_{\rm{e}}(r,\tau)$.~\cite{go.ri.14,bord.07,bo.ga.09} Its charge analogue is defined as~\cite{footnote5} 
\begin{align}
C(r,\tau) &= \sum\limits_{\sigma\sigma'}\langle \hat{n}_{0\sigma} \hat{n}_{r\sigma'}\rangle(\tau)\,\mbox{.}
\label{eq:C}
\end{align}
Correlation functions without a time argument, $S(r)$ and $C(r)$, refer to the ground state of the equilibrium system \eq{eq:HSIAM}, i.e. an impurity coupled to the free electron reservoir. Steady-state correlation functions are indicated with $\tau \rightarrow \infty$. Later we will show that in this limit the time-dependent correlation functions converge to the equilibrium correlations: $S(r,\tau \rightarrow \infty) = S(r)$, as expected from the fact that the quench is intensive. An intuitive measure which quantifies how much of the singlet correlations is contained inside a distance $r$ is the integrated spin correlation function
\begin{align}
\Sigma(r,\tau) &= \sum\limits_{r'=0}^r S(r',\tau)\,\mbox{.}
\label{eq:Sigma}
\end{align}

As discussed below and in \tcite{gu.hi.87,ho.hm.09} the screening length $\xi_k$ can be extracted from $\Sigma(r,\tau)$, by defining it as the length-scale at which a certain fraction $f$ (here we use $f=95\%$) of the correlation lies inside a given distance, i.e.
\begin{align}
\Sigma(\xi_k,\tau)&=(1-f)\Sigma(0,\tau)\,\mbox{.}
\label{sigmaf}
\end{align}

\section{Method}\label{sec:Method}
\begin{figure*}
\includegraphics[width=0.95\textwidth]{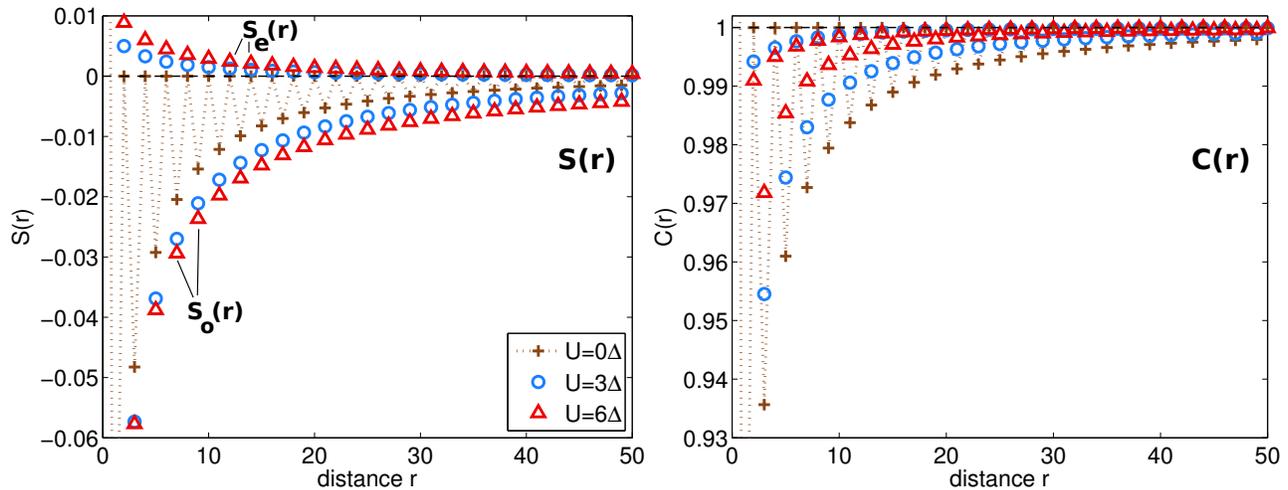}
\caption{(Color online) \emph{Equilibrium correlation functions.} The equilibrium correlation functions for spin- $S(r)$ \textbf{(left)}, \eq{eq:S} and charge $C(r)$ \textbf{(right)}, \eq{eq:C} are shown for short distances from the impurity $r$. The correlation functions at even (odd) distances $S_{\rm{e}}(r)$ ($S_{\rm{o}}(r)$) are indicated explicitly in the left panel. The key is valid for both panels: data for the noninteracting system (brown/pluses) are compared to data for interaction strengths of $U=3\,\Delta$ (blue/circles) and $U=6\,\Delta$ (red/triangles). The alternating behaviour, present for all data sets, is indicated by the dotted line in the $U=0$ data.}
\label{fig:fig1}
\end{figure*}
Here we outline how the correlation functions \eq{eq:S} and \eq{eq:C} are evaluated.
We start with a short discussion of the noninteracting system at equilibrium. In this case we find
\begin{align}
  S^{U=0}(r) &=  \langle\hat{m}_0\rangle \langle \hat{m}_r\rangle  \label{eq:S0rMAG}\\
\nonumber&+ \frac{1}{4}\sum\limits_\sigma \langle 
\cdag{0\sigma} \cnodag{r\sigma} \rangle \left(3\delta_{0r}-\langle 
\cdag{r\sigma} \cnodag{0\sigma} \rangle-2\langle 
\cdag{r\thickbar{\sigma}} \cnodag{0\thickbar{\sigma}} 
\rangle\right)\\
& =\frac{3}{2} \langle 
\cdag{0} \cnodag{r} \rangle\left(\delta_{0r}-\langle 
\cdag{r} \cnodag{0} \rangle\right)\,\mbox{,}\label{eq:S0r} 
\end{align}
where $\langle \hat{m}_r\rangle = \frac{1}{2}\langle \hat{n}_{r\uparrow}-\hat{n}_{r\downarrow}\rangle$, and  the last result holds for the unpolarized case. Here, $c_{r}^\dag/c_{r}$ denote operators for any one of the spin directions $\sigma=\{\uparrow,\downarrow\}$. The opposite spin direction is denoted $\thickbar{\sigma} = -\sigma$. For $U=0$ at equilibrium~\cite{gu.hi.87}
\begin{align}
 C^{U=0}(r) = S^{U=0}(r) + \sum\limits_{\sigma}\langle \hat{n}_{r\sigma} \rangle\,\mbox{.}\label{eq:C0r} 
\end{align}
In the particle-hole symmetric and non spin polarized case the asymptotic limits can be analytically evaluated using results of Ghosh \etal{} in \tcite{go.ri.14} to be
\begin{align}
|S^{U=0}(r)|=\frac{3}{\pi^2}\frac{\Delta}{v_{\text{F}}}
\begin{cases}
(r\frac{\Delta}{v_{\text{F}}})^{-2}&\mbox{ for }r\frac{\Delta}{v_{\text{F}}} \rightarrow \infty\\
 \left[\gamma + \text{ln}\left(r\frac{\Delta}{v_{\text{F}}}\right)\right]^2&\mbox{ for }r\frac{\Delta}{v_{\text{F}}} \rightarrow 0^+\,\mbox{,}
\end{cases}
\label{eq:SU0limits}
\end{align}
for odd $r$ with $\gamma\approx 0.577216$ the Euler-Mascheroni constant. The correlation function becomes zero for even distances $r$. The behaviour of the spin correlation function exhibits a crossover at distance  $\xi^{U=0}\approx\frac{v_{F}}{\Delta}$, which defines a length scale in the noninteracting system.

\begin{figure}
\includegraphics[width=0.42\textwidth]{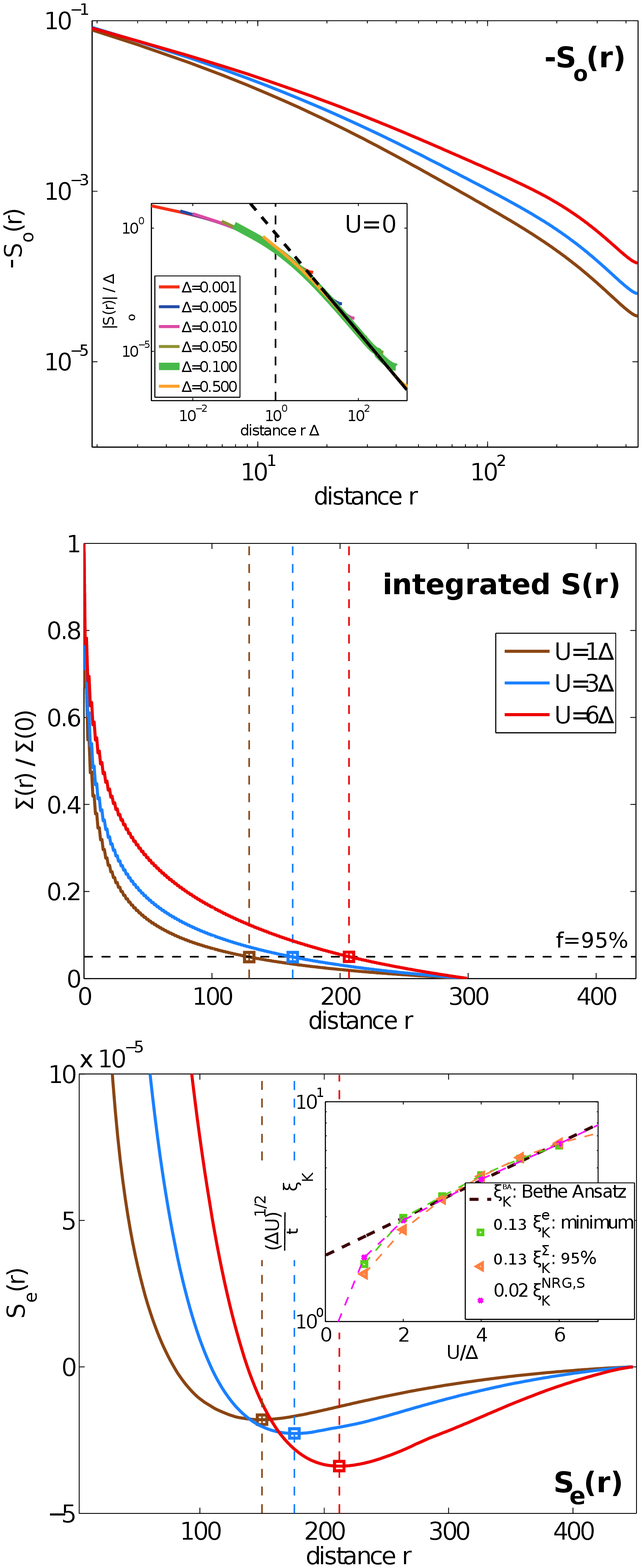}
\caption{(Color online) \emph{Extraction of the equilibrium screening length $\xi_{K}$.} Results in the main panels are shown for $U=1\,\Delta$ (brown)  $U=3\,\Delta$ (blue) and $U=6\,\Delta$ (red).
\textbf{Top:} Spin correlation function for odd distances $S_{\rm{o}}(r)$ (solid line) which displays a  crossover between two different behaviours at small and large $r$. This is  particularly obvious in the $U=0$ results, shown in the inset. Here, $S^{U=0}(r)$ displays the asymptotic behaviour given in \eq{eq:SU0limits}. The large $r$ behaviour is shown by a black dashed line. Our interacting MPS calculations are determined for $\Delta = 0.1$, which corresponds to the thick green line in this plot for $U=0$.
\textbf{Middle:} Integrated correlation function $\Sigma(r)$ of \eq{eq:Sigma}. Dashed vertical lines indicate the distances $\xi_{\text{K}}^{\Sigma}$ inside which $95\%$ of the singlet cloud is contained, which we use to estimate the screening length.
\textbf{Bottom:} Spin correlation function for even distances, $S_{\rm{e}}(r)$. The position $\xi_{\text{K}}^e$ of the minima (circles and vertical dashed lines) is used as alternative definition for $\xi_{K}$. The inset shows these $\xi_{\text{K}}^e$ (green squares) and $\xi_{\text{K}}^\Sigma$ (orange triangles). As reference data we show the BA result and data obtained from an NRG calculation,~\cite{zitk.13} see text.
}
\label{fig:fig2}
\end{figure}

We obtain both $S(r,\tau)$ and $C(r,\tau)$ for zero temperature from computer simulations using MPS~\cite{scho.11} techniques. First, to study ground state correlations, we employ the DMRG~\cite{whit.92,whit.93,scho.05} on a system of length $L$, which is typically $\leq 500$ sites. Second, to study the dynamic formation of the Kondo singlet, we start from a decoupled system in the state \hbox{$\ket{\Psi}=\ket{\uparrow}_{\text{impurity}}\otimes\ket{\text{FS}}_{\text{reservoir}}$}, with a non spin polarized half-filled Fermi sea, at time $\tau=0$ and then switch on the tunnelling term $t'=0.3162\,t$ for times $\tau>0$. The evolution in real-time  is obtained from TEBD.~\cite{vida.04}

MPS based time evolution has proven to be a highly accurate method to evaluate the properties of one-dimensional strongly interacting quantum systems out of equilibrium.~\cite{sa.sc.08,hm.fe.09,schmi.04,ha.fe.06,hm.go.10,znid.11,je.zn.11,hm.ma.09,ki.fu.08,ki.ue.10,ca.ma.02,lu.xi.03,ca.ma.03} The combination DMRG and TEBD is quasi exact as long as the quantum entanglement stays tractable. It has been shown that the main limitation arises due to the growth of entanglement after the quench,~\cite{nu.ga.13, hm.fe.09} which ultimately restricts the available simulation time. Furthermore, since we are interested in the physics resulting from an infinite bath, the maximum available simulation time is restricted by reflections at the lattice boundary and therefore by the finite spatial extent of the system. We have been able to reliably evolve the system long enough to reach a local steady state for all presented data sets. We have checked the convergence of our correlation functions carefully by i) making comparisons to exact data in the $U=0$ system, ii) systematically studying the dependence on the TEBD matrix dimension $\chi$ for finite $U$ and iii) carefully analysing the entanglement entropy. These analyses as well as details of the numerical approach and parameters are provided in \app~\ref{app:NumericalDetails}.

\section{Equilibrium}\label{sec:equilibrium}
We start our discussion by presenting the equilibrium spin ($S(r)$) and charge ($C(r)$) correlation functions. $S(r)$ was first studied by Iishi~\cite{iish.78} and $C(r)$ by Gr{\"u}ner \etal~\cite{me.gr.72,zl.gr.77} who determined the basic spatial dependence and properties. Seminal QMC data~\cite{gu.hi.87} have been extended with the use of the NRG~\cite{bord.07,bo.ga.09} and recently also DMRG.~\cite{ho.hm.09} Here we summarize the most important findings, relevant for the subsequent discussion and we provide details specific to the finite size model and numerical method used. In particular we identify a length-scale in the equilibrium spin correlation function and show later that our nonequilibrium correlation functions converge to the equilibrium correlation function for long times $\tau$.

As shown in \fig{fig:fig1}, both $S(r)$ and $C(r)$ exhibit an oscillating behaviour $\propto \sin{(k_F r)}$. Since the system is half-filled, the Fermi wavevector is $k_F=\frac{\pi}{2}$ and the oscillation period is $r=2$ sites. We first discuss the spin correlations for $U=0$ using \eq{eq:S0r}. In this case we find $S^{U=0}(0) = \frac{3}{2}\langle\hat{n}_{0\uparrow}\rangle(1-\langle\hat{n}_{0\uparrow}\rangle)=\frac{3}{8}$. Furthermore $S^{U=0}(r)$ vanishes for even distances $r$ which follows from general properties of tight binding fermions.~\cite{li.lo.93} The  odd-$r$ correlations $S_{\rm{o}}(r)$ are negative and therefore antiferromagnetic with respect to the impurity. For $U>0$,  $S_{\rm{o}}(r)$ stays negative and increases in magnitude.~\cite{footnote1} At the same time,  the spin correlation function for even distances $S_{\rm{e}}(r)$ gradually develops ferromagnetic correlations at short distances, while it is antiferromagnetic at longer distances. On the one hand, it is the antiferromagnetic component which reflects the screening cloud and signals the formation of the singlet ground state. On the other hand, the ferromagnetic component can be attributed to Coulomb repulsion of opposite spins.~\cite{gu.hi.87} Neither the period nor the phase of the oscillations is changed by the presence of interactions.~\cite{gu.hi.87}

The charge correlation for $U=0$ is linked to the spin correlation via \eq{eq:C0r}. There is oscillatory behaviour between even and odd sites. For even sites the correlation function is unity, while for odd sites it increases monotonically towards unity. For finite interaction strengths we observe a suppression of these Friedel-like oscillations~\cite{fuld.13} with increasing $U$.~\cite{footnote6} At even distances the charge correlations show behavior similar to that of the odd channel, however of a smaller magnitude. The suppression due to the interaction can be traced back to the change in the impurity spectral weight, which develops a narrow Kondo resonance with a width proportional to $T_{\text{K}}$ at the Fermi energy.~\cite{me.gr.72,zl.gr.77} 

While at $U=0$ the characteristic length-scale is $\xi^{U=0}\propto\frac{v_{F}}{\Delta}$, for finite $U$, long range correlations develop which change the behaviour at a distance $\xi_{\text{K}}\propto\frac{v_{\text{F}}}{T_{\text{K}}}$. This crossover characterizing the size of the Kondo spin compensation cloud is visible in the spin correlation function $S(r)$. \Fig{fig:fig2} (top) shows that the antiferromagnetic spin compensation is visible in the spin correlation function at odd distances, $S_{\rm{o}}(r)$. $S_{\rm{o}}(r)$ changes from a logarithmic dependence at small $r\frac{\Delta}{v_{\text{F}}}$ to a power law behaviour at large $r\frac{\Delta}{v_{\text{F}}}$, see \eq{eq:SU0limits}.~\cite{go.ri.14,footnote9} We note that this is different from the Kondo model, where the behaviour is $S(r)\propto r^{-d}$ for $r<\xi_{\text{K}}$ to $S(r)\propto r^{-(d+1)}$ for $r>\xi_{\text{K}}$, with $d$ being the dimensionality of the conduction electron reservoir.~\cite{bord.07,bo.ga.09}

The crossover is difficult to extract directly from numerical data for $S_{\rm{o}}(r)$ since very large system sizes  and small $\Delta$ are required to reach the low $r\frac{\Delta}{v_{\text{F}}}$ limit. We nevertheless found two ways to obtain an estimate for the crossover scale. First, a screening length-scale can be extracted from the integrated correlation function $\Sigma(r)$, see \fig{fig:fig2} (middle). Similarly to \tcites{gu.hi.87,ho.hm.09}, here we denote $\xi_{\text{K}}^\Sigma$ the distance at which $95\%$ of the singlet correlations are covered, i.e. by \eq{sigmaf}. Second, we extract a length scale $\xi_{\text{K}}^{\rm{e}}$ from the spin correlation function at even distances $S_{\rm{e}}(r)$ which for finite $U$ contains both a ferromagnetic component at short distances and the decaying antiferromagnetic one at large distances. As shown in \fig{fig:fig2} (bottom) the function $S_{\rm{e}}(r)$ displays a zero and a minimum and is fit well by a Morse-potential.~\cite{ha.wo.04} We take the position of the minimum as a measure for the crossover scale $\xi_{\text{K}}^{\rm{e}}$. The numerical results obtained with these two crossover scales agree very well and they also agree qualitatively with that obtained by locating the  crossover length between a $r^{-1.4}$ and a $r^{-(1+1.4)}$ behaviour in the $S_{\rm{o}}(r)$ data, which can be estimated from \fig{fig:fig2} (top).

In the inset in \fig{fig:fig2} (bottom) we show that our two estimates, $\xi_{\text{K}}^\Sigma$ and $\xi_{\text{K}}^e$, agree well with established results for the equilibrium screening length. An analytical result, $\xi_{\text{K}}^{\text{BA}}$ (\eq{eq:xiKBA}), 
for the screening length is available via its relation to the Kondo temperature which can be obtained from the BA in the wide band limit by calculating the static spin susceptibility, \eq{eq:xiKBA}. A second benchmark is provided by accurate numerical data from the NRG~\cite{zitk.13,zitk.09} where $T_{\text{K}}^{NRG,\mathcal{S}}$ is defined as the temperature at which the impurity entropy reaches $\mathcal{S}=\frac{\text{ln(2)}}{2}$.~\cite{footnote4} However, while the large $U$ behaviour is universal for all these definitions, the small-$U$ expression, as well as the overall coefficient depends on the specific observable from which it is extracted (spin susceptibility, entropy, etc.). Our data, $\xi_{\text{K}}^\Sigma$ and $\xi_{\text{K}}^e$, agree well with the NRG result $\xi_{\text{K}}^{NRG,\mathcal{S}}$, they are all compatible with a simple exponential growth in $U$
\begin{align}
 \xi_{\text{K}}&\propto e^{\frac{\pi}{16 \Delta}U}
\label{eq:xiKnum}
\end{align}
For $U>2\,\Delta$ this agrees with the BA prediction \eq{eq:xiKBA} which features an additional factor of $\sqrt{\Delta U}$. We note that for $U\leq2\,\Delta$ no well defined local moment has formed~\cite{ho.hm.09} i.e. $U$ is too small for the system to develop a pronounced local moment regime in between the low- and the high-temperature limit. Our data also compare very well with those presented in an extensive study of length-scales in the SIAM on a finite lattice at equilibrium in \tcite{ho.hm.09}.

These results indicate that the method presented here is reliable in producing unbiased correlation functions at equilibrium which exhibit the characteristic features of a Kondo screening cloud. The cloud is well contained within the numerically tractable lattice size $L\le 500$ for $U\leq6\,\Delta$, see \app~\ref{app:NumericalDetails}. Therefore, we focus our calculations on $U\leq 6\,\Delta$ when discussing the time-dependent correlation functions.

\begin{figure*} 
\includegraphics[width=0.95\textwidth]{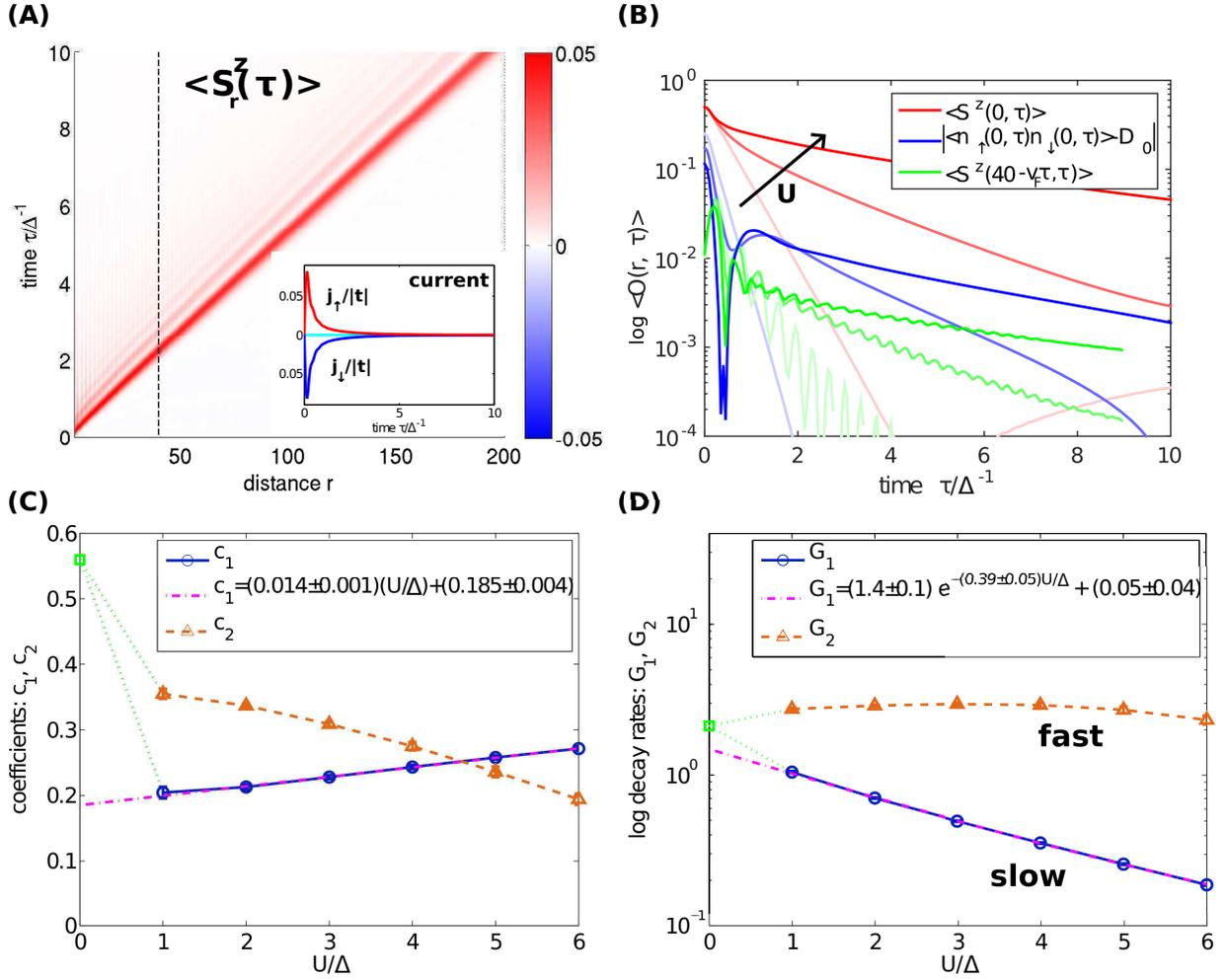}
\caption{(Color online) \emph{Time evolution of local expectation values.} \textbf{(A)} Evolution of the spin-z density $\langle S^z_r (\tau)\rangle$ as a function of distance the $r$ and time $\tau$. We plot data for $U=3\,\Delta$ and the color axis are cut off below the maximum for better visibility of fine structures. The vertical black line indicates a cut at distance $r=40$. The inset shows the time dependent spin-current~\cite{nu.ga.13} $\langle j_\sigma(\tau)\rangle =i\,\pi\,t'\langle f_{\sigma}^{\dag}  c_{1\sigma}- c_{1\sigma}^{\dag}  f_{\sigma} \rangle$ at the bond next to the impurity: $j_\uparrow(\tau)$ (red) and $j_\downarrow(\tau)$ (blue) and the total current $j(\tau)$ (cyan). \textbf{(B)} Local spin-z density at the impurity site $r=0$ and at $r=40$ as a function of time. Note that the $r=40$ data are shifted such that the light cone coincides with the origin. From the local double occupancy at the impurity site $r=0$ we subtracted the equilibrium values $D_0(U) = \{0.25, 0.1745, 0.1153\}$ for $U=\{0,3,6\}\,\Delta$ as obtained by DMRG. All data are plotted for three interaction strengths $U=\{0,3,6\}\,\Delta$,from lighter to darker color as indicated by the black arrow. \textbf{(C)} Fit coefficients $c_{1/2}$ of the double exponential fit of the decay of $\langle S^z(r,\tau)\rangle$ to its equilibrium value as a function of $U$ (cf. \eq{eq:fit}). The magenta line indicates a linear fit to the coefficient of the slow component $c_1$. \textbf{(D)} Decay rates $G_{1/2}$ of the double exponential fit of the decay of $S^z$ to its equilibrium value as a function of $U$. The magenta line indicates an exponential fit to the decay rate of the slow component $G_1$. The single-exponential behaviour at $U=0\,\Delta$ is indicated in green.}
\label{fig:figA5}
\end{figure*}
\begin{figure*}
\includegraphics[width=0.95\textwidth]{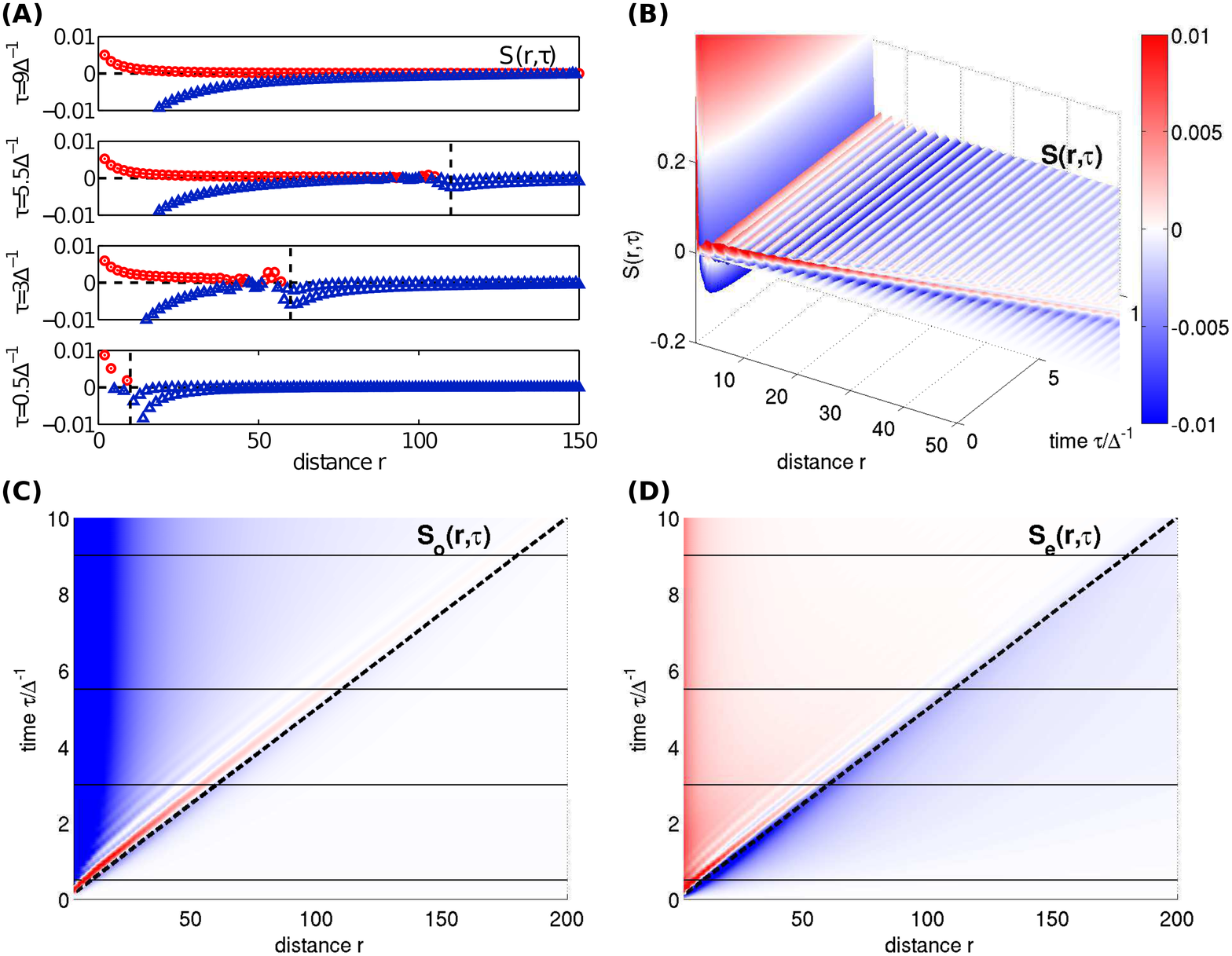}
\caption{(Color online) \emph{Overall profile of the space-time evolution of the spin correlation function $S(r,\tau)$, \eq{eq:S}.} \textbf{(A):} $S(r,\tau)$ as a function of distance $r$ for four different times: $\tau=0.5\,\Delta^{-1}$, $\tau=3\,\Delta^{-1}$, $\tau=5.5\,\Delta^{-1}$ and $\tau=9\,\Delta^{-1}$, from bottom to top. The ferromagnetic contribution is shown as red circles while antiferromagnetic components are displayed as blue triangles. The horizontal black dashed lines indicate the zeros. Far behind the signal wake the antiferromagnetic component coincides with $S_{\rm{o}}(r,\tau)$ and the ferromagnetic component with $S_{\rm{e}}(r,\tau)$. The signal front (light cone), travelling with speed $v_{\text{F}}\approx2t$, is indicated as a vertical black dashed line. In all panels the signal at very small distances which is of order unity has been cut off for better visibility. \textbf{(B):} Three dimensional visualization of $S(r,\tau)$. The colorbar of (B) is also valid for panels (C) and (D). \textbf{(C):} Spin correlation function at odd distances: $|S_{\rm{o}}(r,\tau)|$ (top view). \textbf{(D):} Spin correlation function at even distances: $|S_{\rm{e}}(r,\tau)|$ (top view). In both bottom panels horizontal black lines indicate times at which data are presented in panel (A). The light cone defined by $v_{\text{F}}$ is marked by a thick black dashed line. All data shown are for $U=3\,\Delta$.}
\label{fig:fig7}
\end{figure*}

\section{Time evolution of local observables}\label{sec:localObs}
Before beginning the discussion of the temporal evolution of spatial correlations we present the time evolution of the local observables with a focus on the impurity site.

At time $\tau=0$ we start with a spin-$\uparrow$ particle at the impurity and a non spin polarized half-filled Fermi sea: \hbox{$\ket{\Psi}=\ket{\uparrow}_{\text{impurity}}\otimes\ket{\text{FS}}_{\text{reservoir}}$}. For the connected equilibrium system in the thermodynamic limit we expect a uniform and non spin polarized density that is $\langle n_{0\uparrow}\rangle (\tau\rightarrow\infty) = 0.5$, $\langle n_{0\downarrow}\rangle (\infty) = 0.5$, $\langle n_0\rangle (\infty) = 1$ and $\langle S^z_0\rangle (\infty) = 0$. The impurity double occupation in a non-interacting system or in the high temperature limit is $\langle n_{0\uparrow}n_{0\downarrow}\rangle^{U=0}(\infty)=0.25$.~\cite{di.sc.13} For finite interaction strength the evolution is non trivial.

\Fig{fig:figA5} (A) shows the expectation values of the spin-z projection $\langle S^z_r\rangle (\tau) = \frac{1}{2}\left(\langle n_{r\uparrow}\rangle(\tau)-\langle n_{r\downarrow}\rangle(\tau)\right)$. Due to particle-hole symmetry, the total charge density $\langle n_{r} \rangle(\tau)$ is unity.

Indeed we find that, following the hybridization quench, the excess spin-$\uparrow$ on the impurity is transported away. This happens essentially with the Fermi velocity $v_{\text{F}}\approx2t$ as shown by the major signal in \Fig{fig:figA5}(A).

The resulting missing spin-$\uparrow$ density is exactly compensated by the spin-$\downarrow$ density due to particle-hole symmetry. This compensation takes place simultaneously and completely symmetrically in both spin channels as is evident from the spin-$\uparrow$ and spin-$\downarrow$ currents shown in the inset in \Fig{fig:figA5}(A). The time integral over the spin-current reveals that half a particle is transferred in or out of the impurity in a time on the order of $\approx 3 \Delta^{-1}$ for $U=3\,\Delta$.

\Fig{fig:figA5} (B) shows the local evolution of expectation values as a function of time and interaction strength. All expectation values converge to their respective, exactly known, equilibrium values as noted above. The time-evolved double occupancy also converges to the equilibrium results obtained by DMRG. This indicates that our time evolution is accurate and unbiased, at least for large times. For more convergence checks and uncertainty estimates we refer the interested reader to \app~\ref{app:NumericalDetails}.

At a certain distance $r$ from the impurity, a resulting signal arrives at $\tau\approx\frac{r}{v_{\text{F}}}$. This signal is oscillating and strongly damped in time see \Fig{fig:figA5}(B). With increasing interaction strength $U$, the initial spike becomes dampened in amplitude, but the oscillating tail gains in weight. The signal at $r=40$ in the double occupancy  has the same structure 
on a scale of $10^{-3}$ around its equilibrium value.

In the following we consider the temporal decay of the spin-z density at the impurity in detail. Previous studies using the time-dependent NRG for the SIAM~\cite{an.sc.05} and analytical calculations at the Toulouse point of the anisotropic Kondo model~\cite{lo.ke.05} found that the initial dynamics of $\langle S^z_o\rangle(\tau)$ is governed by a fast time scale $\propto\frac{1}{\Delta}$ while the eventual relaxation exhibits a long time scale $\propto\frac{1}{T_{\text{K}}}$ governed by Kondo physics. These results were confirmed by bold-line quantum Monte Carlo simulations~\cite{co.gu.13} on the SIAM which observed these two time scales collapsing into one for an applied bias voltage.

From our data we find that, as expected, the spin-z density at $U=0$ decays in a single-exponential manner
\begin{align*}
 \langle S^{z,U=0}(r=0,\tau)\rangle&=(0.561\pm0.001)e^{-2(1.060\pm0.002) \tau \Delta}\\
&+(0.0001\pm0.0001)\,\mbox{,}
\end{align*}
hence it features the fast hopping time scale $\mathcal{T}_{U=0}\propto \frac{1}{\Delta}$. For finite $U$, a double exponential decay develops
\begin{align}
\label{eq:fit}
 \langle S^{z}(r=0,\tau)\rangle&=c_1 e^{-G_1\tau \Delta}+c_2 e^{-G_2\tau \Delta}\,\mbox{.}
\end{align}
In panel (C, D) we show the results of this data analysis with respect to the interaction strength $U$ in the available range of $U\in[1,6]$. We identify one fast, exponential decay $G_2\approx 2(1.4\pm0.2)$ yielding a $U$ independent time scale $\mathcal{T}_{\text{fast}}\propto \frac{1}{\Delta}$ similar to $\mathcal{T}_{U=0}$. The corresponding coefficient $c_2$ decreases in magnitude with increasing $U$. In contrast, the more interesting slow exponential decay $G_1$ has a coefficient $c_1$ which becomes more and more prominent with increasing $U$. In particular the coefficient $c_1$ exhibits a linear behaviour in $U$
\begin{align*}
 c_1(U)&=(0.014\pm0.01)\frac{U}{\Delta}+(0.185\pm0.004)\,\mbox{.}
\end{align*}
The slow decay rate $G_1$ is exponentially small in $U$
\begin{align*}
 G_1(U)&=(1.4\pm0.1)e^{-2(0.19\pm0.02)\frac{U}{\Delta}}+(0.05\pm0.04)\,\mbox{.}
\end{align*}
This implies that the Kondo physics manifests itself in the local dynamic observable $\langle S^z_0\rangle(\tau)$ in the form of a slow time constant $\mathcal{T}_{\text{slow}}\propto e^{2(0.19\pm0.02)\frac{U}{\Delta}}$ which 
has the same $U$ behaviour as the  Kondo temperature (cf. \eq{eq:xiKnum}).

The double occupancy $\langle n_{0\uparrow}n_{0\downarrow}\rangle(\tau)$ converges to its equilibrium value with the same dominant slow decay (within numerical uncertainty) as observed in the spin-z density for finite $U$. At $U=0$ the fast decay rate is twice rate observed in the spin-z density at $U=0$.

Performing the same analysis for distances $r$ away from the impurity that considers $\langle S^z_r\rangle(\tau)$ we again observe the same decay as at the impurity site within the fit uncertainty, see \Fig{fig:figA5} (B). This supports the quasi particle picture introduced in \tcite{me.ho.13} which translates the physics at the impurity via emission of spin dependent quasi particles to a given distance $r$.

\begin{figure*}
\includegraphics[width=0.95\textwidth]{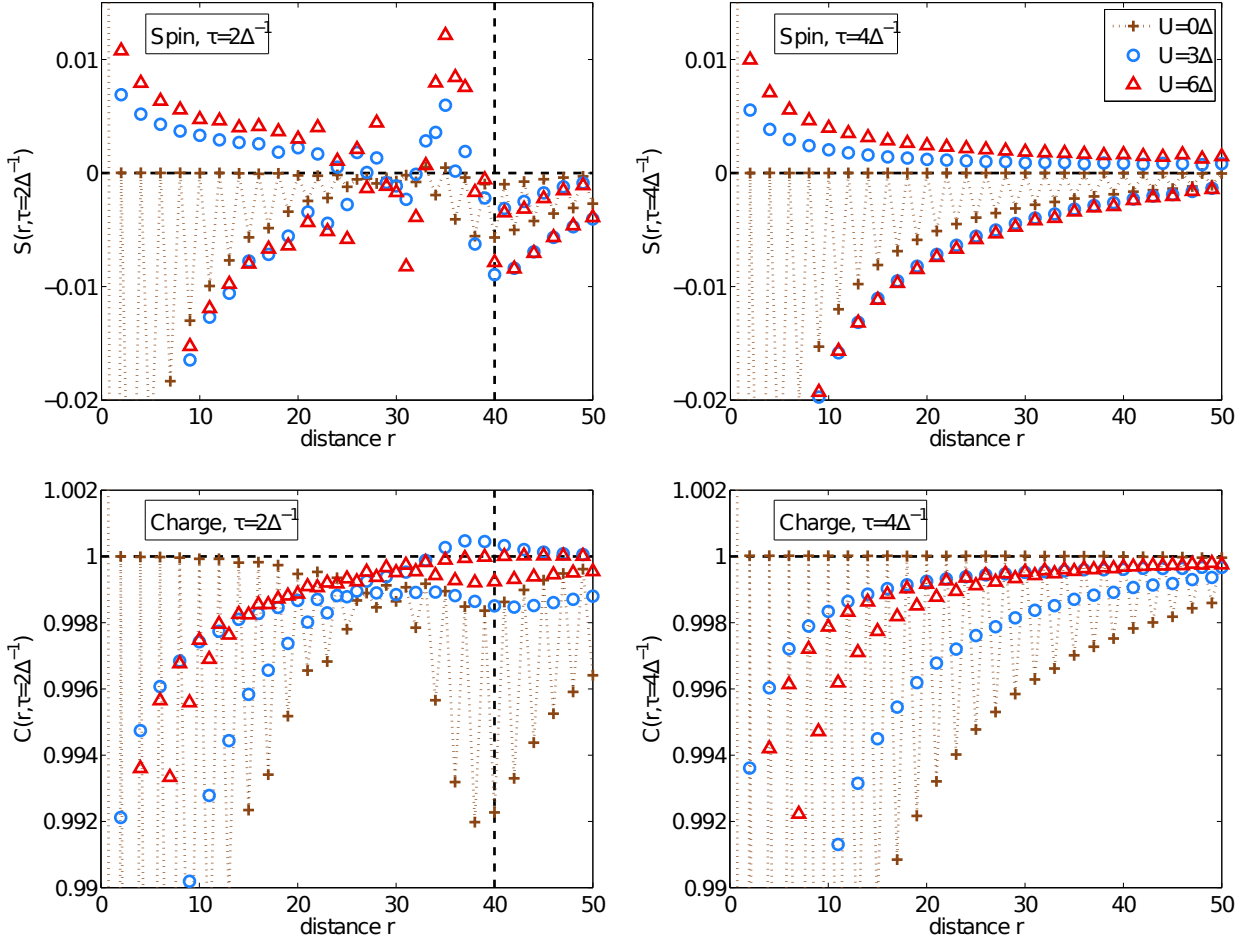}
\caption{(Color online) \emph{Detail of the time evolved correlation functions} The \textbf{top (bottom)} panel shows $S(r,\tau)$ ($C(r,\tau)$) for $\tau=2\,\Delta^{-1}$ \textbf{(left column)} and $\tau=4\,\Delta^{-1}$ \textbf{(right column)}. The signal front, travelling at speed $v_{\text{F}}\approx2t$ is indicated as a vertical dashed black line. Results are shown for different values of $U$ as indicated in the key which is valid for all panels. The alternating behaviour is indicated by a dotted line in the $U=0$ data.}
\label{fig:fig4}
\end{figure*}

\section{Time evolution of the screening cloud}\label{sec:nonequilibrium}
A very interesting question is how the spatial structure of the Kondo screening cloud develops, i.e. whether and how it is approached in a nonequilibrium time evolution starting from an initial state without Kondo physics. The question was recently first studied in pioneering work on the case of an exactly solvable model, namely the anisotropic Kondo model at the Toulouse point.~\cite{hoff.12,me.ho.13,footnote2} A complementary numeric study using the time-dependent NRG~\cite{le.an.14} was performed shortly afterward on the isotropic Kondo model extending and confirming the analytical results from the Toulouse limit.

Let us now investigate the formation of spatial correlations $S(r,\tau)$, \eq{eq:S} and $C(r,\tau)$, \eq{eq:C} after switching on the tunnelling between the Anderson impurity and the reservoir electrons. We first focus on the major characteristics of $S(r,\tau)$, displayed in \Fig{fig:fig7}. At time $\tau=0$ the initial configuration is a product state $\ket{\Psi(\tau=0)}=\ket{\uparrow}_{\text{impurity}}\otimes\ket{\text{FS}}_{\text{reservoir}}$. Using \eq{eq:S0rMAG}, we find at the impurity: $S^{U=0}(r=0,\tau=0) = \frac{3}{4}\langle\hat{n}_0\rangle-\frac{3}{2}\langle\hat{n}_{0\uparrow}\rangle\langle\hat{n}_{0\downarrow}\rangle= 0.75$, where $\hat{n}_0=\sum\limits_\sigma\hat{n}_{0\sigma}$ since we have $\langle\hat{n}_{0\uparrow}\rangle = 1$ and $\langle\hat{n}_{0\downarrow}\rangle = 0$.

After the quench in the hybridization $t'$, we observe a strong signal in $S(r,\tau)$, travelling at the Fermi velocity $v_{\text{F}}\approx 2t$ which defines a light cone. It has been attributed to the propagation of quasi particles in \tcite{me.ho.13}. The propagating signal front divides the space-time into two regions: i) a region at large times and small distances where the correlation function is directly affected by the impurity and Kondo correlations develop and ii) a region at small times and large distances where small structures from the quench are observed. In \se~\ref{sec:inside} and \se~\ref{sec:outside} we discuss the detailed behaviour of the correlation functions within these two regions. The signal front itself carries a large chaotic disturbance in its wake and a small monotonic perturbation ahead of it. 

As discussed below in detail, the time evolved data $S(r,\tau)$ converges to the equilibrium correlation functions $S(r)$ within the light cone. Already a look at the almost vertical structures in \Fig{fig:fig7} (C, D) for times $\tau\geq8\,\Delta^{-1}$ and a comparison of the line plots for $\tau=6.5\,\Delta^{-1}$ and $\tau=9\,\Delta^{-1}$ for small distances $r$ hint at the convergence to a local steady state within the light cone.

\Fig{fig:fig7} (D) reveals that, as expected from the equilibrium state, a ferromagnetic correlation develops for even distances $r$ in $S_{\rm{e}}(r,\tau)$ within the light cone for finite $U$, while outside the light cone this correlation function is always antiferromagnetic. As shown in \fig{fig:fig7} (A, C), the wake behind the light cone carries a ferromagnetic signal also at odd distances $r$ that is in the otherwise antiferromagnetic $S_{\rm{o}}(r,\tau)$ for all $U$. We interpret this signal as remnant information of the spin which occupied the impurity at $\tau=0$ before the quench. Following the signal wake, all characteristic features of the equilibrium correlation function $S(r)$ develop quickly on a qualitative level. Far behind the signal wake the antiferromagnetic component coincides with $S_{\rm{o}}(r,\tau)$ and the ferromagnetic component with $S_{\rm{e}}(r,\tau)$.

A closer look, as provided in \fig{fig:fig4}, reveals that the nonequilibrium correlation functions gradually develop the characteristic features of the equilibrium correlation functions $S(r)$ and $C(r)$ for $r < v_{F} \tau$. As a precursor of the equilibrium structure, the spin correlation function $S(r,\tau)$ develops the oscillatory behaviour of its equilibrium counterpart inside the light cone. That is, it oscillates from an antiferromagnetic correlation at odd distances $r$ to a ferromagnetic correlation at even $r$ for finite $U$ or to zero at $U=0$. This structure of the phase and period of these oscillations in space is fixed over time inside the light cone. However, the light cone induces a phase shift of $\pi$ in the nodal structure of the correlation function. We attribute this phase shift to the initial state of the Fermi sea. It takes place across the broad signal behind the light cone visible at around $r\approx30$ in the figure. As a function of $U$ the same behaviour is present inside the light cone as at equilibrium, apart from the chaotic disturbance at the light cone. The qualitative functional form of the correlation functions develops quickly in the wake of the light cone. However, its amplitude overshoots the expected equilibrium value slightly and then gradually decays to it at a much slower time scale, see discussion in \se~\ref{sec:inside}.

The charge correlation function $C(r,\tau)$ gradually develops reduced Friedel-like oscillations with increasing $U$, as observed at equilibrium. We find $C(r,\tau)<1$ except at distances $r<3$ and in the vicinity of the signal front.

In the following we investigate in detail the interplay of characteristic time- and length-scales and their dependence on the interaction strength.

\begin{figure*}
\includegraphics[width=0.95\textwidth]{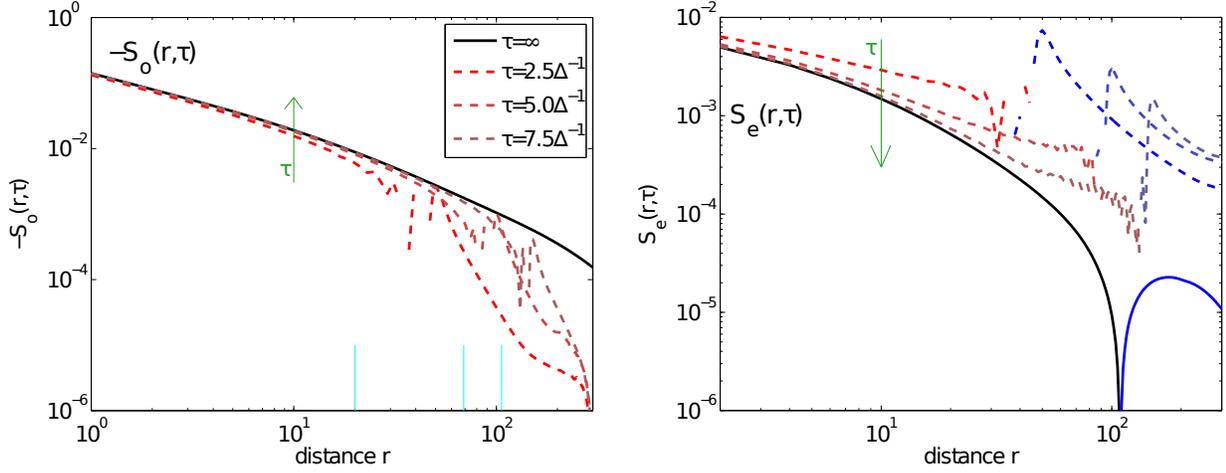}
\caption{(Color online) \emph{Convergence of the nonequilibrium data to the equilibrium results inside the light cone.} The spin correlation functions ($S(r,\tau)$ at odd ($S_{\rm{o}}(r,\tau)$ \textbf{left}) and even $S_{\rm{e}}(r,\tau)$ \textbf{right}) distances are depicted as a function of the distance $r$ for three different times: $\tau=2.5\,\Delta^{-1}$, $\tau=5\,\Delta^{-1}$ and $\tau=7.5\,\Delta^{-1}$ in a log-log fashion (dashed lines). We plot $-S_{\rm{o}}(r,\tau)$ since it is almost entirely negative while $S_{\rm{e}}(r,\tau)$ is positive inside the light cone and negative outside, see \fig{fig:fig7}. Blue lines represent $|S_{\rm{e}}(r,\tau)|$ in regions where $S_{\rm{e}}(r,\tau)$ is negative. The key depicted in the left panel is valid for both panels. Green arrows mark the direction of increasing time $\tau$. Data from the equilibrium simulation are plotted in solid black and referred to as $\tau=\infty$ in the key. The vertical cyan lines in the left panel mark those distances at which cuts through the data as a function of $\tau$ are presented in \fig{fig:fig6}. All data shown are for $U=3\,\Delta$.}
\label{fig:fig5}
\end{figure*}
\begin{figure*} 
\includegraphics[width=0.95\textwidth]{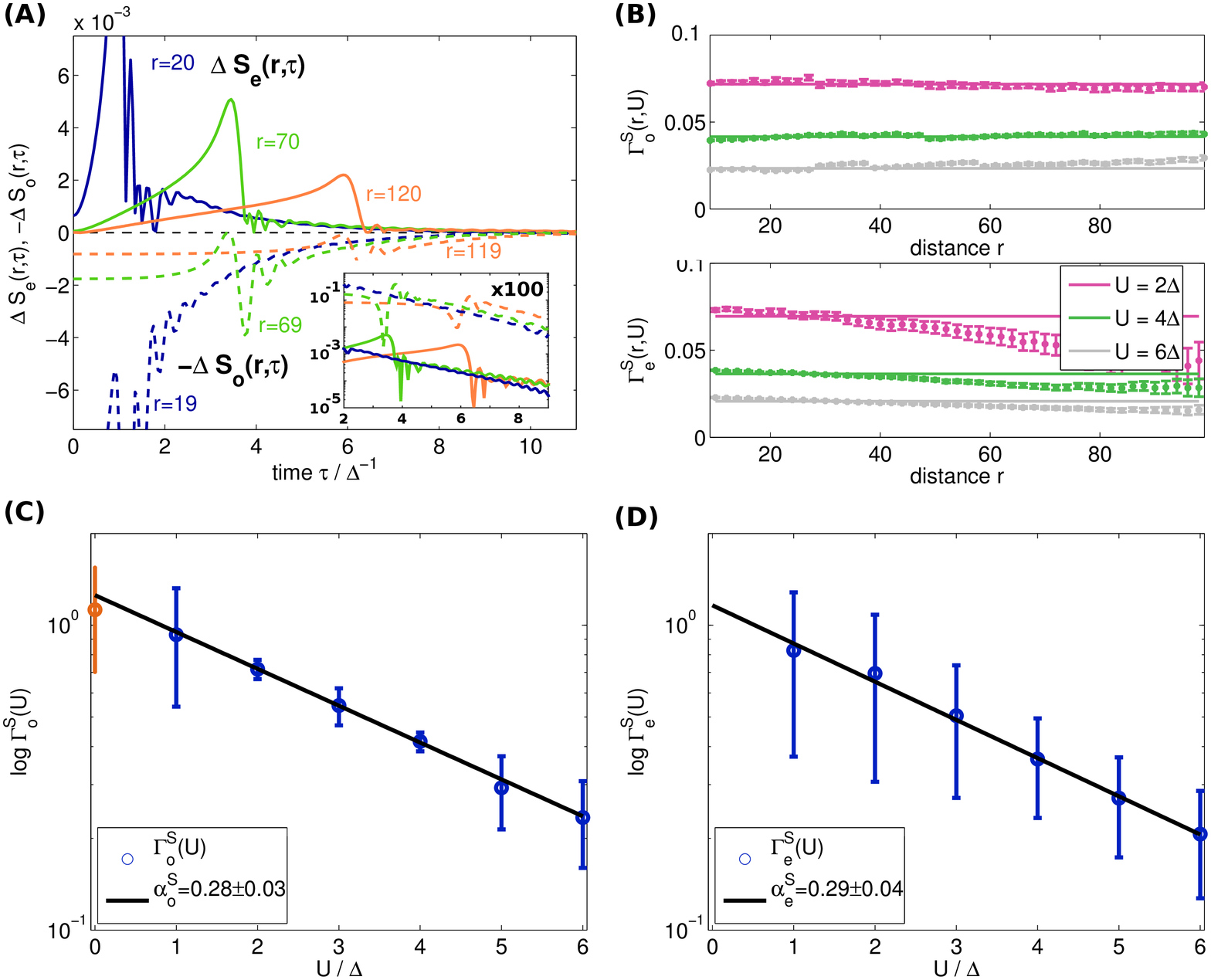}
\caption{(Color online) \emph{Identification of a dynamic time scale in the exponential convergence to equilibrium inside the light cone.} \textbf{(A)} Deviation of the time dependent spin correlation function  from the equilibrium spin correlation function at even (solid) and odd (dashed) distances $\Delta S_{\rm{e/o}}(r,\tau)$ as a function of time $\tau$ for three different distances: $r=20/19$ (blue), $r=70/69$ (green) and $r=120/119$ (orange). For better visibility we show $-\Delta S_{\rm{o}}(r,\tau)$. The signal changes behaviour at the light cone at $\tau\approx \frac{r}{v_{\text{F}}}$ which is visible as a large hump. The region inside the light cone is to the right of this hump. The semilogarithmic zoom to this region in the inset shows an exponential convergence. To separate the curves of the even and the odd component in the inset, data for odd distances are multiplied by a factor of $100$. Data shown are for $U=3\,\Delta$. \textbf{(B)} Extracted decay rates $\Gamma^S_{\rm{e/o}}(r,U)$ as a function of distance $r$ for three values of interaction strength $U=\{2,4,6\}\,\Delta$. \textbf{(C, D)} Spatially averaged exponential decays at odd (C) and even (D) distances $\Gamma^S_{\rm{o/e}}(U)$ as a function of the interaction strength $U$ (circles). The dynamic scale $\alpha^S_{\rm{o}}=0.28\pm0.03$ and $\alpha^S_{\rm{e}}=0.29\pm0.04$, \eq{eq:Sdecay2}, (solid black) is extracted via a single-exponential fit of the respective data where only data points for $U\geq1\Delta$ are considered (blue circles) and data for $U<1\Delta$ are excluded (orange circles). For details of the fits see \app~\ref{app:fitDetail}.}
\label{fig:fig6}
\end{figure*}

\subsection{Inside the light cone}\label{sec:inside}
Next we discuss the spin correlation function $S(r,\tau)$ inside the light cone. \Fig{fig:fig5} shows the convergence of $S_{\rm{o}}(r,\tau)$ and $S_{\rm{e}}(r,\tau)$ to their equilibrium $S_{\rm{o}}(r)$ and $S_{\rm{e}}(r)$ values for large times in detail. For large times the odd component is antiferromagnetic while the even component exhibits a sign change from ferromagnetic at small distances to antiferromagnetic at large distances (blue curves) as discussed in the equilibrium results. The vanishing ferromagnetic component represents a related measure for the extent of a screening cloud.~\cite{gu.hi.87}

In the following we identify a time scale at which large correlations with the impurity develop inside the light cone, i.e. for distances $r\leq v_{\text{F}}\tau$, see \fig{fig:fig0b}. In \fig{fig:fig6} (left) we show the difference between the time-dependent spin correlation function and the equilibrium result: $\Delta S_{\rm{o/e}}(r,\tau) = |S_{\rm{o/e}}(r,\tau)-S_{\rm{o/e}}(r)|$. This quantity exhibits contributions from the travelling signal, which show up in the form of large spikes at times $\tau\approx\frac{r}{v_\text{F}}$. We first focus on the convergence in time at fixed distances $r$. For times beyond the signal wake $\tau\propto \frac{r}{v_{\text{F}}}$, the qualitative structure of correlations has established itself, i.e. Kondo correlations have reached the given distance $r$. We find that soon after the signal wake $S(r,\tau)$ converges to the equilibrium result exponentially in time
\begin{align*}
 \Delta S_{\rm{o/e}}(r,\tau) &\propto e^{-\tau \Delta\cdot\Gamma^S_{\rm{o/e}}(r,U)}\,\mbox{,}
\end{align*}
see \fig{fig:fig6} (A, inset). Note that this implies that the curves move ``as a whole". We determine $\Gamma^S_{\rm{o/e}}$ by a single-exponential fit in time of $\Delta S_{\rm{o/e}}(r,\tau)$, successively for distances $r\in[40,120]$. We observe that $\Gamma^S_{\rm{o/e}}(r, U)$ is only weakly dependent on $r$, with odd distances $r$ being especially stable, see \fig{fig:fig6} (B), while $\Gamma^S_{\rm{e}}$ has larger uncertainties and some drift at large $r$. The uncertainty increases slightly with distance $r$ which is also due to the smaller available fit intervals in $\tau$. A two exponential decay as in ~\se~\ref{sec:localObs} featuring also a fast time scale $\propto\frac{1}{\Delta}$ and independent of $U$ might be present in the data, but cannot be identified due to the presence of the signal at the light cone which overshadows this fast decay. In general the fit quality improves with increasing $U$. Details about the data analysis and uncertainty estimates are provided in \app~\ref{app:fitDetail}.

In order to condense this information we consider a mean value
\begin{align*} 
\Gamma^S_{\rm{o/e}}(U) := \frac{1}{N_r}\sum_{r=40(41)}^{120}\Gamma^S_{\rm{o/e}}(r,U)\mbox{,}
\end{align*}
with $N_r$ the number of distances in the respective odd or even interval, see \fig{fig:fig6} (B).

Our first main result is that $\Gamma^S_{\rm{o/e}}(U)$ shows an exponential behaviour in $U$, like the Kondo scale, \eq{eq:xiKnum}
\begin{equation}
\boxed{
\begin{aligned}
\Gamma^S_{\rm{o/e}}(U) \propto e^{-\alpha^S_{\rm{o/e}} \frac{U}{\Delta}}\,\mbox{.}
\end{aligned}
}
\label{eq:Sdecay2}
\end{equation}
\Fig{fig:fig6} (C, D) show the fit in $U$ to \eq{eq:Sdecay2} where we find $\alpha^S_{\rm{o}}=(0.28\pm0.03)$ and $\alpha^S_{\rm{e}}=(0.29 \pm 0.04)$ which is similar to the BA result in the wide band limit for the Kondo temperature, $T_{\text{K}}^{\text{BA}}\propto e^{-\alpha_{\text{BA}}\frac{U}{\Delta}}$, $\alpha_{\text{BA}}=0.196$, compare \eq{eq:xiKBA}. The deviation of the effective exponent $\alpha^S_{\rm{o/e}}$ from $\alpha_{\text{BA}}$ may be due to the fact that it is particularly difficult to reach the common asymptotic limit in space and in time for large $U$. Note that $\langle S(0,\tau)\rangle \langle S(r,\tau)\rangle\ll S(r,\tau)$ thus the connected correlation function displays essentially the same behaviour as $S(r,\tau)$.

We conclude that the formation of Kondo correlations inside the light cone is a process which involves two major time scales. The first time scale is fast and determined by the lattice Fermi velocity $v_{\text{F}}$, which defines the light cone and develops qualitatively correct correlations in $S(r,\tau)$ and $C(r,\tau)$. The second time scale is slow and depends exponentially on $U$. This process sets in after the qualitatively correct correlations have built up with $v_{\text{F}}$ and renormalizes the correlation functions which then converge at an exponential rate \eq{eq:Sdecay2} $\alpha^S_{\rm{o/e}} \propto T_{\text{K}}$ to the equilibrium result.

The SIAM is related to its low energy realization, the antiferromagnetic, symmetric Kondo model via the Schrieffer Wolf transformation,~\cite{sc.wo.66} which effectively integrates out charge fluctuations. The two models share common features in their low energy behaviour, most prominently the Kondo scale $T_{\text{K}}$. Note however, that the correlation functions of the two models have a very different spatial structure in general. It is therefore interesting to compare our results to recently obtained ones for the Kondo model. In \tcite{le.an.14} Lechtenberg \etal{}, studied a coupling quench in the symmetric Kondo model using the time-dependent NRG as well as second order perturbation theory. Similarly to our results for the SIAM, they found that in the Kondo model spin correlations develop rather rapidly on the scale of the Fermi velocity. In the linear response to a magnetic field, at equilibrium they observed another, slower time scale similar to $\frac{1}{T_{\text{K}}}$. Our results unambiguously and quantitatively identify this common slower scale $\frac{1}{T_{\text{K}}}$ beyond linear response, directly from the nonequilibrium time evolution of correlation functions.

Charge correlations at equilibrium do not exhibit Kondo physics. We observe that the charge time dependent correlation functions $C(r,\tau)$ do exhibit qualitatively the same convergence to equilibrium as the spin correlations $S(r,\tau)$ that is with a time constant exponentially large in $U$ (not shown). The same analysis as for the spin using \eq{eq:Sdecay2} yields respective coefficients for the charge correlation function $\alpha^C_{\rm{o/e}}\approx(0.3\pm0.1)$. That is, the exponent is the same as for the spin albeit with a larger uncertainty. We attribute this to the resolution of the spin in the correlators present in $C(r,\tau)$. Note that this is true neither for the local density which does not show such a scale nor for the mean-field result $C^{\text{mf}}(r,\tau)\propto1$.

\begin{figure*}
\includegraphics[width=0.95\textwidth]{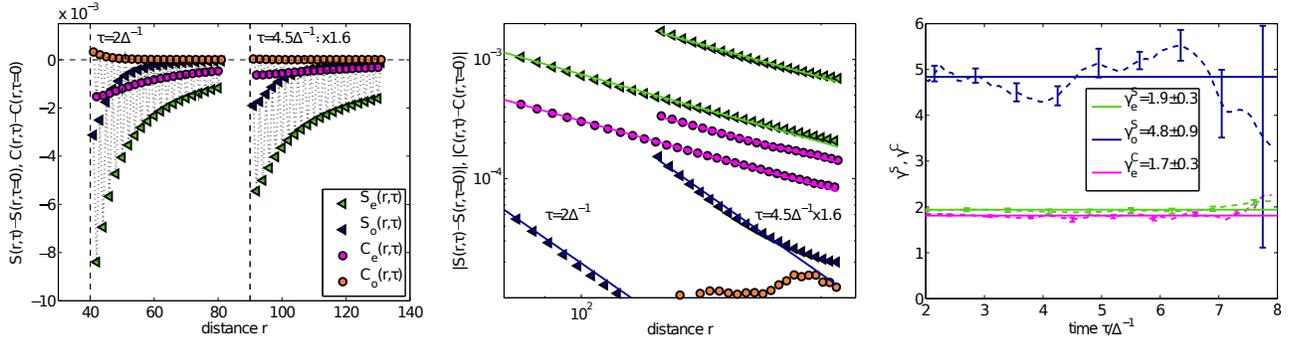}
\caption{(Color online) \emph{Correlations outside the light cone.} \textbf{Left:} $S(r,\tau)$ (blue/green triangles) and $C(r,\tau)$ (orange/magenta circles) outside the light cone with their $\tau=0$ values subtracted and resolved by even and odd distances $r$. Data are depicted as a function of $r$ for two times: $\tau=2\,\Delta^{-1}$ and $\tau=4.5\,\Delta^{-1}$. Note that we only display data in the vicinity of the light cone (vertical dashed black lines). For reasons of representation only, data  at $\tau=4.5\,\Delta^{-1}$ are scaled by a factor of $1.6$. \textbf{Middle:} Modulus of the data in the left panel plotted in a double logarithmic representation. The key of the left panel is also valid. Straight lines represent power law fits. \textbf{Right:} The extracted power law coefficients $\gamma^S_{\rm{e/o}}$ (dashed blue/green) and $\gamma^C_{\rm{e}}$ (dashed orange) are plotted as a function of time $\tau$. Error bars, shown for some data points only, are estimated from the non-linear fit presented in the middle panel and its susceptibility to changing the fit intervals. Solid horizontal lines indicate weighted time and interaction strength averages: $\gamma^S_{\rm{e}}=1.9\pm0.3$, $\gamma^S_{\rm{o}}=4.8\pm0.9$ and $\gamma^C_{\rm{e}}=1.7\pm0.3$. Data in all three panels are for $U=3\,\Delta$.}
\label{fig:fig10}
\end{figure*}

\subsection{Outside the light cone}\label{sec:outside}
For distances $r> v_{\text{F}} \tau$, i.e. outside the light cone, see \fig{fig:fig0b}, we find decaying correlation functions $S(r,\tau)$ and $C(r,\tau)$ as a function of $r$, see \fig{fig:fig10}. As before, both spin and charge correlation function show alternating behaviour from site to site. The overall magnitude of both correlation functions decreases over time and the charge correlation function is of a smaller magnitude than the spin correlation function for all except very early times. To identify the correlations generated by the quench, we subtract the initial correlation $S(r,\tau=0)$ and $C(r,\tau=0)$ from the time-dependent data.

The second main result of this work is that correlations outside the light cone are power-law suppressed
\begin{equation}
\boxed{
\begin{aligned}
 |S(r,\tau)-S(r,0)| &\propto r^{-\gamma^S_{\rm{o/e}}} \\
 |C(r,\tau)-C(r,0)| &\propto r^{-\gamma^C_{\rm{e}}}\,\mbox{,}
\end{aligned}
}
\label{eq:PowerOutside}
\end{equation}
with slightly time-dependent exponents $\gamma^S_{\rm{o/e}}$ and $\gamma^C_{\rm{e}}$. Due to the finite size of the system, we only have a limited set of data available to extract the asymptotic decay outside the light cone. We start the extraction of power law exponents at distances $r_s=v_{\text{F}} \tau+35$ to avoid spurious contributions from the light cone and end it at $r_e=L-70$ to avoid a bias originating from the boundary at $L=450$. From the separate fits for odd/even distances we obtain $\gamma^S_{\rm{o}}\approx1.9\pm0.3$ and $\gamma^S_{\rm{e}}\approx4.8\pm0.9$. The charge correlation function exhibits a power law decay $\gamma^C_{\rm{e}}\approx1.7\pm0.3$ for the odd component, while the even component's behaviour cannot be identified within our numerical accuracy due to the small magnitude of the correlations. The fit has been performed in the same fashion as presented in \App~\ref{app:fitDetail} but here we estimate the uncertainty in the $\gamma$'s from the fluctuations of the respective $\gamma$ upon changing the start ($r_s$) and endpoint ($r_e$) of the fit. Within this uncertainty, the values are independent of $U$ and $\tau$.

Considering the fact that extracting exponents from numerical data is challenging, our results agree quite well with two recent studies of similar models exhibiting comparable low energy physics. First, in \tcite{me.ho.13} Medvedyeva \etal{} obtained time dependent correlation functions at the Toulouse point of the anisotropic Kondo model with a linear dispersion. In an analytic calculation in several limits, neglecting Friedel oscillations, they showed that outside the light cone the commutator spin-z correlation function $\langle\left[\hat{S}^z_0 \hat{S}^z_r(\tau)\right]_-\rangle$, which is related to the linear response to a perturbation, vanishes. For the anti-commutator, which is proportional to our $S(r,\tau)$ (see \eq{eq:S}), however, they obtained a power law decay $r^{-2}$ at zero temperature (see equation 27 in their work). They found the initial entanglement in the reservoir Fermi sea to be responsible for the power law decay of the anti-commutator correlation function.

Moreover, second order perturbation theory results (Lechtenberg \etal{}, \tcite{le.an.14}) suggest that initial correlations of the Fermi sea transfer to the time dependent correlations outside the light cone. Here again a $r^{-2}$ power law decay outside the light cone was obtained, this time for the isotropic Kondo model with antiferromagnetic coupling $J$. Our study of the symmetric SIAM finds an $r^{-2}$ decay for $S_{\rm{o}}(r,\tau)$ outside the light cone, which we attribute to the same structures of the electronic reservoir in the three studies.
We are not aware of any previous reports of even-distance decay exponents $\gamma^S_{\rm{e}}\propto r^{-5}$.

\section{Conclusions}\label{sec:conclusions}
We studied the time-dependent formation of the spin screening cloud in the single impurity Anderson model. Starting from an unentangled state we switched on the impurity-reservoir hybridization and followed the subsequent dynamics in real time. We used the density matrix renormalization group to obtain ground states and time evolving block decimation to obtain spin and charge correlation functions. From these correlation functions we obtained characteristic time and length-scales. Our results agree with previous calculations at equilibrium and for local observables out of equilibrium. We found that the nonequilibrium correlation functions converge to the equilibrium results for long times. 

In the time-dependent data, we identified a linear spreading of signals travelling at the lattice Fermi velocity which has been referred to as a light cone in recent literature on the build-up of a screening cloud at the Toulouse point of the anisotropic Kondo model.~\cite{hoff.12,me.ho.13} We observed a ferromagnetic response in the wake of the signal at the light cone. While \tcite{hoff.12,me.ho.13} identified the Kondo temperature as an inverse time scale in the anisotropic Kondo model outside the light cone, for the symmetric Kondo model it was observed as a time scale in an equilibrium linear response calculation to a magnetic perturbation following an initial fast decay.~\cite{le.an.14} We observe directly from the nonequilibrium time evolution of correlation functions that in the SIAM too, the structure of the correlation functions inside the light cone emerges on two time scales. The qualitative core of the correlation functions develops rapidly, at the lattice Fermi velocity. This includes the phase and period of oscillations as well as fixed ferromagnetic and antiferromagnetic domains. These correlations then reach their equilibrium values  exponentially slowly in time, defining a dynamic rate which shows the same exponential $U$-dependence as the Kondo temperature.

Outside the light cone, we find a power-law decay of the correlation functions $\propto r^{-\gamma^{S/C}_{\rm{o/e}}}$,  with essentially interaction strength- and time-independent exponents, \eq{eq:PowerOutside}. In addition to the $r^{-2}$ decay also observed in the Kondo model~\cite{hoff.12,me.ho.13,le.an.14} we find a decay $\propto r^{-5}$.

Our results could be experimentally verified in one-dimensional optical lattices featuring two fermionic species. By monitoring the evolution of the spin correlations in time, our findings provide the basis for extracting information about the dynamic scale and therefore, indirectly about the Kondo screening cloud dynamics as well as the system parameters.

Possible future extensions to this work include the study of the inverse process. Starting from a coupled impurity-reservoir system and investigating the Kondo destruction after switching the hybridization to zero would allow study of the time reversed situation. It would also be very interesting to study the effects of a bias voltage on the Kondo screening process using a two-terminal setup as in \tcite{nu.ga.13}. Further interesting extensions involve the study of conduction bands with singularities or testing of implications of the nonequilibrium fluctuation-dissipation theorem. Also, calculations away from particle-hole symmetry or with applied magnetic fields are feasible.

\begin{acknowledgments}
We gratefully acknowledge fruitful discussion with Sabine Andergassen, Masud Haque, Fabian Heidrich-Meisner, Kerstin T. Oppelt and Shreyoshi Ghosh. We thank Rok \v{Z}itko for providing the NRG Ljubljana code.~\cite{zitk.13} This work was partly supported by Austrian Science Fund (FWF) P24081-N16, and SFB-ViCoM projects F04103 and F04104 as well as NaWi Graz. MN thanks the Forschungszentrum J{\"u}lich, in particular the Autumn School for Correlated Electrons, for hospitality.
\end{acknowledgments}

\appendix
\setcounter{figure}{0}  \renewcommand{\thefigure}{A\arabic{figure}}

\section{Numerical Details}\label{app:NumericalDetails}
\begin{figure*}
\includegraphics[width=1.0\textwidth]{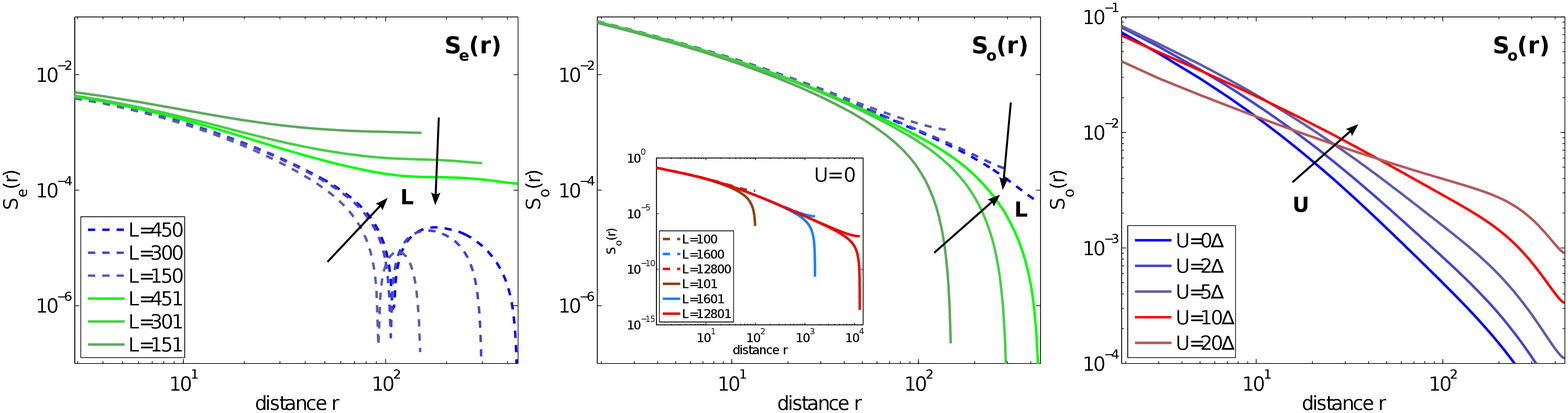}
\caption{(Color online) \emph{Finite size analysis and the imposed limits on interaction strength.} Finite size effects in the equilibrium spin correlation function at even ($S_{\rm{e}}(r)$, \textbf{(left)}) and  at odd ($S_{\rm{o}}(r)$, \textbf{(middle)}) distances. For each one, we compare even total lattice sizes $L=\{150,300,450\}$ (blue) to odd $L=\{151,301,451\}$ (green). The key of the left panel is valid for both, the left and the middle panel. Both main panels show data for $U=3\,\Delta$. The inset in the middle panel depicts the evolution with $U$ for a total system size of $L=450$. The \textbf{right} panel shows the evolution of $S_{\rm{o}}(r)$ with increasing interaction strength $U=\{0,2,5,10,20\}\,\Delta$ for $L=450$. The correlation becomes qualitatively wrong if $U$ is too large for a given $L$.}
\label{fig:fig11}
\end{figure*}
\begin{figure*}
\includegraphics[width=0.95\textwidth]{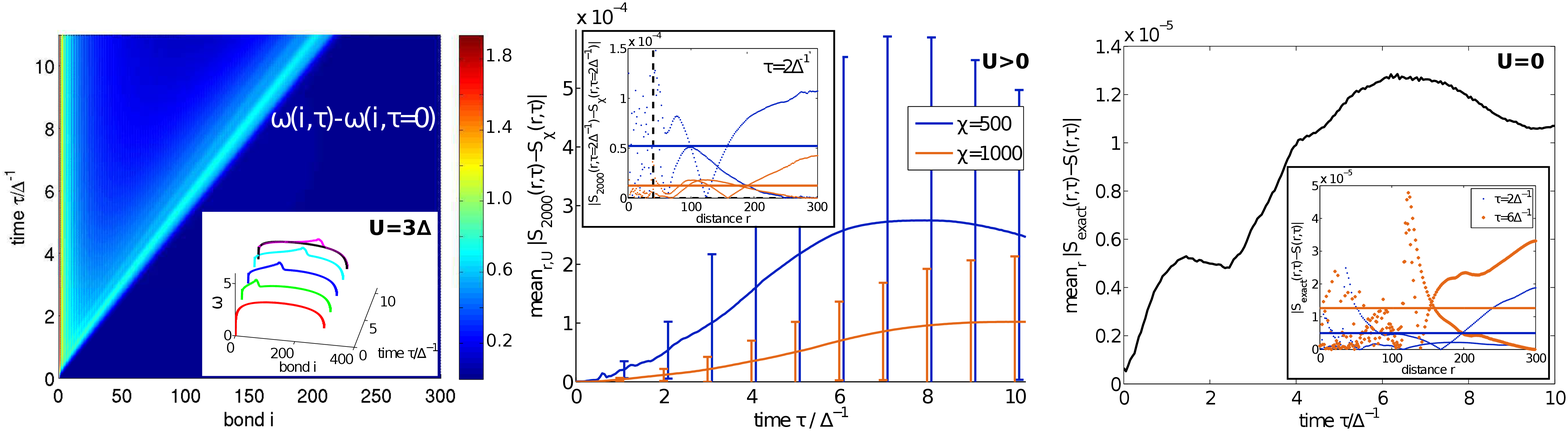}
\caption{(Color online) \emph{Quality of the DMRG and TEBD data.} \textbf{Left:} Bond and time resolved entanglement entropy $\omega(i,\tau)$. In the main panel we subtracted the $\omega(i,\tau=0)$ data, to highlight changes caused by the time evolution. The inset shows cuts through the $\omega(i,\tau)$ raw data at constant times. The black line is the result of a corresponding equilibrium simulation. The area hidden by the inset is homogeneously dark blue which corresponds to $\omega(i,\tau)-\omega(i,\tau=0)\equiv0$. Data shown is for $U=3\,\Delta$. \textbf{Middle:} Convergence of the interacting spin correlation function with increasing TEBD matrix dimension $\chi$. Modulus of the residuals $|S_{2000}(r,\tau)-S_{\chi}(r,\tau)|$, benchmarking the quality of the time evolution with increasing TEBD matrix dimension $\chi$. We show results comparing $\chi=2000$ with $\chi=500$ (blue) and $\chi=2000$ with $\chi=1000$ (orange). In the main panel we show the residuals averaged over distance and interaction strength as a function of time $\tau$. In the inset, the spatially resolved residuals are plotted at time $\tau=2\,\Delta^{-1}$ and for $U=3\,\Delta$. \textbf{Right:} Comparison of the noninteracting spin correlation function as obtained by TEBD $S(r,\tau)$ and the noninteracting spin correlation as obtained exactly $S_{\text{exact}}(r,\tau)$. The main panel again shows the spatially averaged absolute distance while the inset shows spatial resolution for two different times $\tau=\{2, 6\}\,\Delta^{-1}$. Note that the blue (orange) each one belongs to one data set only, which is alternating.}
\label{fig:fig12}
\end{figure*}
In this Appendix we specify details on our numerical analysis carried out via the DMRG~\cite{whit.92,whit.93} and TEBD,~\cite{vida.04} and we present the DMRG and TEBD parameters used. In addition, we discuss finite size effects and the convergence as a function of auxiliary parameters specific to the numerical method applied, as well as the stability of the real time evolution. Our numerical implementation of the DMRG and TEBD is flexible, is parallelized and exploits two Abelian symmetries: particle number $\hat{N}$ and spin projection $\hat{S}^z$. To find ground states we use the two site DMRG algorithm with successive single site DMRG steps. The time evolution is based on a second order Suzuki-Trotter decomposition of the propagator.~\cite{scho.05,scho.11}

After extensive studies of the dependence of our results on auxiliary system parameters we found converged results for a Trotter time step of $\delta\tau =0.05\,t^{-1}$. We used DMRG and TEBD matrix dimensions of $\chi=2000$ states and always started the DMRG optimization from a half-filled system in the canonical ensemble where alternating up and down spins are chosen as the seed. A detailed discussion is available in \tcite{nu.ga.13} in the context of previous work.

\Fig{fig:fig11} shows the equilibrium DMRG calculation of the correlation functions. The influence of the finiteness of the lattice is twofold:

i) The equilibrium spin correlation function $S(r)$ displays an even-odd effect as a function of the total system size $L$: While for even $L$, $S_{\rm{o}}(r)$ converges from above to its $L\to\infty$ value, for odd $L$ it converges from below. $S_{\rm{e}}(r)$ converges in the opposite way. For odd $L$ an extra spin-$\uparrow$ gives a spurious total magnetisation. For the equilibrium simulations, in the main part of the paper, we have chosen $L=450$, since it supports a half-filled and non spin polarized system. The spin correlation function at  $r\leq150$ is converged,  as can be seen in \fig{fig:fig11} by comparing the $L=450$ and the $L=300$ results. Larger distances are influenced by $L$ because $S(r)$ is a non-local quantity.  Nevertheless, even for larger distances, no qualitative differences are observed between the $L=450$ and $L=300$ data. When performing the time evolution we use $L_{\text{equilibrium}}+1$ lattice sites, including the impurity, so that the reservoir is non-magnetized and half-filled. With this choice the correlation functions of the equilibrium and the nonequilibrium simulations become comparable.

ii) The size of the Kondo screening cloud becomes exponentially large in $U$. It is therefore important to identify the characteristics of finite size effects with increasing $U$. In \fig{fig:fig11} (right) we plot data with increasing $U$ for fixed $L$ and study the behaviour of $S_{\rm{o}}(r)$. From $U=0$ to $U=6\,\Delta$ the correlation function follows a monotonic trend and qualitatively the same behaviour. However, the curves for $U=10\,\Delta$ and $U=20\,\Delta$ are qualitatively different. This indicates that these values of $U$ are too large for the given $L$ as expected from the size of $\xi_{\text{K}}^{\text{BA}}$ which becomes of the order of $L\approx200$ sites here, see \eq{eq:xiKBA}. In the present work we therefore restrict ourselves to values of $U\leq6\,\Delta$.

Next we show that our time evolution yields a controlled accuracy using a DMRG/TEBD matrix dimension of $\chi=2000$. The bipartite entanglement~\cite{scho.11} $\omega(i,\tau)=-\text{tr}\left[\hat{\rho}_{L/R}(\tau)\text{ln}(\hat{\rho}_{L/R}(\tau))\right]$ provides an estimate of the time when TEBD becomes unreliable for a fixed $\chi$. This is signalled by a sharp increase in $\omega$. Here $\hat{\rho}_{L/R}$ denotes the reduced density matrix to the left (L) or to the right (R) of a lattice bipartition at bond $i$.  \Fig{fig:fig12} (left) shows the entanglement increase, which turns out to mostly affect the region next to the impurity and the major propagating signal at $r=v_{\text{F}} \tau$. In our simulations we find that $\chi=2000$ is sufficient to account for the additionally generated entanglement which is not much larger than in the equilibrium case. In addition we investigate the direct influence of increasing $\chi$ on the interacting spin correlation function $S_\chi(r,\tau)$ by comparing results using $\chi=2000$ with results obtained at a smaller $\chi$. \Fig{fig:fig12} (middle) shows the modulus of the deviation $|S_{2000}(r,\tau)-S_{\chi}(r,\tau)|$. We calculate this deviation at each point in space $r$ and time $\tau$ and for $U=\{0,1,2,3,4,5,6\}\,\Delta$. The deviation fluctuates over space with systematic signatures at the light cone and beyond it, while the interior of the light cone looks chaotic. The results are almost independent of $U$. We find that the space $r$ and interaction $U$ averaged deviation grows over time and becomes of the order of $\mathcal{O}(5\cdot10^{-4})$ for $\chi=500$ and $\mathcal{O}(1\cdot10^{-4})$ for $\chi=1000$ within the reachable simulation time. Furthermore, for $U=0$ we compare the correlation functions obtained via TEBD with the numerically exact ones  (\eq {eq:S0rMAG}) in \Fig{fig:fig12} (right). As one can see, the maximum deviation occurs at the boundary far from the impurity with a maximum error of $\approx 1\cdot10^{-5}$. 

We conclude that for simulations of non-local correlation functions within the DMRG and TEBD in the SIAM the major limiting factor is the computation time $T\propto L\,(\chi)^3$. This is due to the large matrix dimensions $\chi$ needed to achieve accurate results and is furthermore complicated by the fact that the SIAM exhibits exponentially long correlation lengths which require large lattice sizes $L$. The accuracy can be controlled by benchmarking against exactly known $U=0$ data and for finite $U$ by increasing the TEBD matrix dimension $\chi$. Furthermore all the scales extracted in the main text, $\alpha^S_{\rm{o/e}}$ and $\gamma_{\rm{C/S_{o/e}}}$ are retrieved from two subtracted correlation functions, in which we expect errors to further compensate. 

\section{Extraction of the dynamic energy scale}\label{app:fitDetail}
In the following we provide details of the data analysis of the dynamic scale $\alpha_{o/e}$ as discussed in \se~\ref{sec:inside}, which is valid for both even and odd distances. First, we obtain the time dependence of the spin correlation function by performing a non-linear fit in time $\tau$ to the spin correlation function for fixed distances $r$ and given interaction $U$: $\Delta S(\tau|r,U)$ (see \se~\ref{sec:inside}) using $f(\ve{\phi} = (c_1,\Gamma(r,U)), \tau)=c_1 e^{-\Gamma(r,U) \tau}$ with $2$ fit parameters $\ve{\phi}$. The data are single-exponential plus oscillations and exhibit an eventual systematic bias close to the lattice border and due to the signal front at the light cone. For each $r$ we manually choose intervals $[\tau_s(r,U),\tau_e(r,U)]$ for the fit in time in order to minimize these influences. Typically we choose fit intervals which start $r_s\approx10$ sites behind the light cone and extend up to $r_e\approx 250$ for large $U$. For small $U$ the data become noise before this $r_e$ is reached and we adjust the endpoints accordingly. We estimate the fit uncertainty $\Delta \Gamma(r, U)$ by $\Delta\phi_i\approx \sqrt{C_{ii}}$ where $\ve{C}=(\ve{J}^\dag \ve{J})\eta^2$ is the estimated covariance, $\ve{J}=\frac{\partial f(\ve{\phi},\tau_i)}{\partial \alpha_j}$ is the fit Jacobian and $\eta^2=\frac{\ve{r}^T\ve{r}}{N_\tau(r,U)-p}$ is the mean square error defined by the residuals $r_i=\Delta S(\tau_i|r,U)-f(\ve{\phi},\tau_i)$ on $N_\tau(r,U)$ data points in time $\Delta S(\tau_i|r,U)$. These estimates are consistent with those obtained from fluctuations upon changing $\tau_s(r,U)$ and $\tau_e(r,U)$. Secondly we condense the $r$ dependence by averaging $\Gamma(r,U)$ over distances $r$. We make use of a Bayesian approach with Gaussian error statistics. We obtain the weighted mean value $\Gamma(U) = \frac{1}{P}\sum\limits_r\frac{1}{\widetilde{\Delta\Gamma}(r,U)^2}\Gamma(r_i,U)$ with $P=\sum\limits_r\frac{1}{\widetilde{\Delta\Gamma}(r,U)^2}$ and a weighted error estimate $\Delta \Gamma(U) = \frac{1}{\sqrt{P}}$ where the weights are obtained from $\Delta\Gamma(r,U)$. Third, we obtain the $U$ dependence of the exponent considering data for $\Gamma(U)$ for $N(U)=6$ data points at $U=\{1,2,3,4,5,6\}\,\Delta$. The data $\Gamma(U)$ can be fitted very well by a single-exponential in $U$: $\Gamma(U) = c_2 e^{-\alpha U}$. The same scheme as in the first step is used to estimate the uncertainty $\Delta \Gamma$. We assume correlated data i.e. $\eta^2=\frac{\ve{r}^T\ve{r}}{\widetilde{N}_{\text{eff}}}$, with $\widetilde{N}_{\text{eff}}\approx\frac{N(U)-p}{2 N_{\text{corr}}}\approx\frac{6-2}{2\cdot6}$ which enlarges the uncertainty by a factor of $\sqrt{3}$ compared to the naive value.

\bibliography{kondoScreening,footnotes}{}

\begin{thebibliography}{130}
\bibitem[{foo({\natexlab{a}})}]{footnote3}
\bibinfo{note}{We denote real time by $\tau$.}
\bibitem[{foo({\natexlab{b}})}]{footnote10}
\bibinfo{note}{We denote all operators by a hat: $\hat{O}$, except for
  elementary fermion annihilation / creation operators: $c^\nag / c^\dag$.}
\bibitem[{foo({\natexlab{d}})}]{footnote8}
\bibinfo{note}{In the wide-band limit a featureless (constant) electronic
  density of states of the reservoirs is assumed which has a bandwidth much
  larger than the characteristic energy scales of the system coupled to the
  reservoir. The retarded single particle Green's function is then given by
  $G^R(\omega)=-\frac{1}{2 W} \text{ln}\left(\frac{\omega-W}{\omega+W}\right)$
  where $W$ is the half bandwidth.}
\bibitem[{foo({\natexlab{e}})}]{footnote11}
\bibinfo{note}{Note that although our Hamiltonian is spin rotation invariant,
  in our system $S(r,\tau)\neq 3 \langle \hat{S}^z_0 \hat{S}^z_r\rangle(\tau)$
  since our initial state is not.}
\bibitem[{foo({\natexlab{f}})}]{footnote5}
\bibinfo{note}{Here we also consider the disconnected contribution to the
  charge correlation function in order to compare to \tcite{gu.hi.87}.}
\bibitem[{foo({\natexlab{g}})}]{footnote1}
\bibinfo{note}{A non-monotonic behaviour of the magnitude of the spin
  correlation function $S(r)$ with increasing $U$ has been reported for small
  $U$.~\cite{gu.hi.87} Our data points in $U$ are not dense enough to comment
  on this aspect. Our results suggest a monotonic dependence.}
\bibitem[{foo({\natexlab{h}})}]{footnote6}
\bibinfo{note}{In \tcite{gu.hi.87} it has been observed that $C(r)$ exhibits a
  suppression $r\ll \frac{v_{\text{F}}}{T_{\text{K}}}$, which gradually
  recovers to the full Friedel value at large distances. We cannot comment on
  the full recovery due to our limited system sizes.}
\bibitem[{foo({\natexlab{i}})}]{footnote9}
\bibinfo{note}{While the asymptotic limits of \eq{eq:SU0limits} are obtained
  for $U=0$, it has been argued in \tcite{go.ri.14} that they stay valid in the
  interacting system.}
\bibitem[{foo({\natexlab{j}})}]{footnote4}
\bibinfo{note}{Making use of the NRG Ljubljana code~\cite{zitk.13} we studied
  the model as presented in the main text by calculating the impurity
  thermodynamics. We used a $Z$ discretization scheme, $Nz=7$ points for
  z-averaging and a cutoff $\Lambda=1.8$. The presented results are converged
  in all auxiliary numerical parameters.}
\end{thebibliography}

\end{document}